\documentclass[aps,prl,twocolumn,superscriptaddress,showpacs]{revtex4}

\usepackage{CJK}

\usepackage{graphicx}

\usepackage{dcolumn}

\usepackage{bm}
\usepackage[normalem]{ulem}
\usepackage{color}

\usepackage{epstopdf}

\newcommand{\bq}{\mathbf{q}}

\newcommand{\bv}{\mathbf{v}}

\newcommand{\br}{\mathbf{r}}

\newcommand{\bu}{\mathbf{u}}

\newcommand{\sep}{ \ \ \ , \ \ \ }

\newcommand{\beq}{\begin{equation}}
\newcommand{\eeq}{\end{equation}}
\newcommand{\beqn}{\begin{eqnarray}}
\newcommand{\eeqn}{\end{eqnarray}}
\newcommand{\pp}{\partial}
\newcommand{\dd}{{\rm d}}
\newcommand{\ee}{{\rm e}}

\newcommand{\eqs}{Eqs }
\newcommand{\fig}{Fig.\ }

\newcommand{\cO}{{\cal O}}

\newcommand{\g}{\gamma }

\newcommand{\la}{\langle}

\newcommand{\ra}{\rangle}

\begin{document}

\begin{CJK*}{UTF8}{gbsn}
%\preprint{APS/123-QED}

\title{Squeezed in three dimensions, moving in  two: Hydrodynamic theory of 3D incompressible easy-plane polar active fluids
%Birds, magnets, soap, and sandblasting: surprising connections %in the theory to incompressible polar active fluids in 2D
}
%\author{Leiming Chen}
\author{Leiming Chen 
	(陈雷鸣)}
\email{leiming@cumt.edu.cn}
\affiliation{School of Physical science and Technology, China University of Mining and Technology, Xuzhou Jiangsu, 221116, P. R. China}
\author{Chiu Fan Lee}
\email{c.lee@imperial.ac.uk}
\affiliation{Department of Bioengineering, Imperial College London, South Kensington Campus, London SW7 2AZ, U.K.}
\author{John Toner}
\email{jjt@uoregon.edu}
\affiliation{Department of Physics and Institute of Theoretical
	Science, University of Oregon, Eugene, OR $97403$}

\begin{abstract}
We study the hydrodynamic behavior of three dimensional (3D)
incompressible collections of self-propelled entities in contact with a
momentum sink in a state with non-zero average velocity,
hereafter called 3D easy-plane incompressible polar active fluids.
We show that the hydrodynamic model for this system belongs to the
same universality class as that of an equilibrium system, namely a special
3D anisotropic magnet. The latter can be further mapped onto yet another
equilibrium system, a DNA-lipid mixture in the sliding columnar phase.
Through these connections we find a divergent renormalization of the
damping coefficients in  3D easy-plane incompressible polar active fluids,
and  obtain  their equal-time velocity correlation functions.

\end{abstract}
\maketitle
\end{CJK*}

\begin{figure}
	\begin{center}
		\includegraphics[scale=.3]{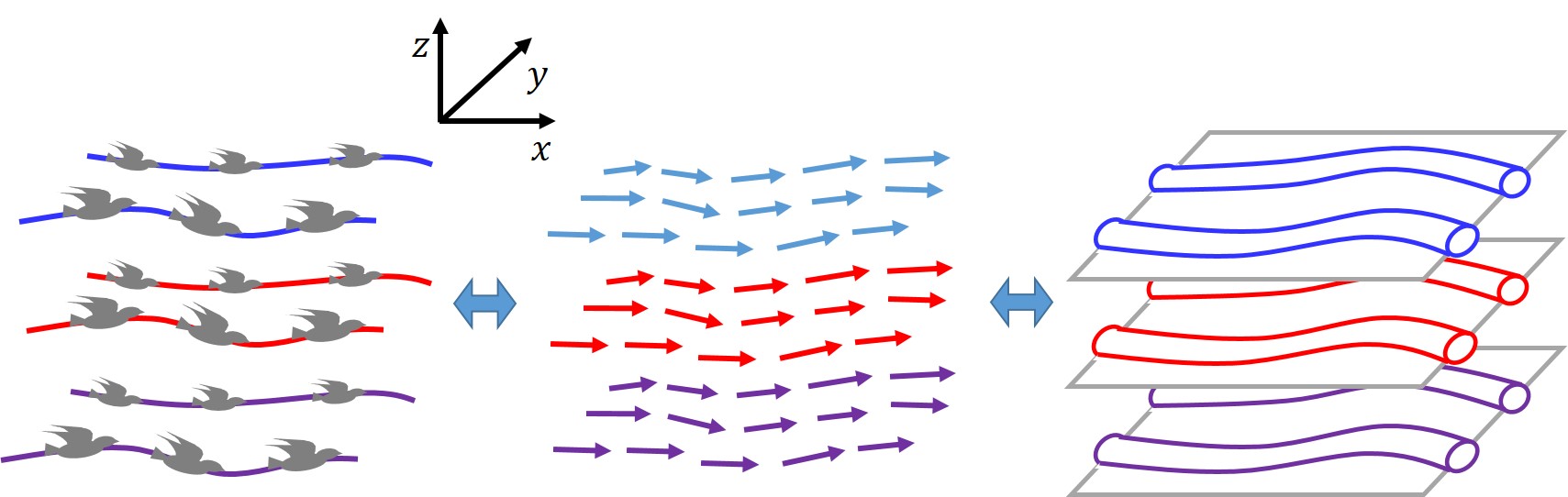}
	\end{center}
	\caption{
			A schematic illustrating that  an incompressible flock with an  easy-plane of flying,   an easy-plane magnet, and a sliding columnar phase have identical equal-time statistics in the hydrodynamic limit: the fluctuations of the flow lines, magnetic lines, and the columns share the same scaling behavior.
			}
	\label{fig:cartoon}
\end{figure}

Diverse distinct systems can share identical large-distance, scale invariant
properties. This ``universality", which occurs when the distinct systems
share common symmetries, can link seemingly disparate
areas of physics. In this paper, we reveal such a surprising connection
between two such areas: Active matter \cite{Active}  and self-assembly
of biomimetic material. Specifically,
we demonstrate that a particular three dimensional (3D)  phase of
self-propelled, incompressible agents - namely, what we call an
``incompressible easy-plane active fluid" can  exhibit precisely  the same
scaling behavior as an equilibrium mixture of DNA and cationic lipids in
a hypothesized phase known as the ``sliding columnar"
phase  \cite{Column1,Column2,Column} (\fig \ref{fig:cartoon}).

%%%%%%%%%%%%%%%%%%%%
%%%%%%%%%%%%%%%%%%%%%%%%

An ``incompressible easy-plane active fluid" is a collection of
self-propelled entities, which could be, e.g., living creatures like bacteria,
or synthetic self-propelled objects like Janus particles \cite{Janus} or
``Quinke rotators" \cite{Bartolo}. The term ``active" refers in this context to
their self-propulsion. ``Incompressible" means that we consider specifically
systems in which these self-propelled particles move in such a way that
they do not change their density, either because they are tightly packed,
so that no space is available for them to change their density, or because
of long-ranged interactions, like the ions in a plasma. By ``easy-plane",
we mean systems in which the motion of these entities is preferentially
parallel to some plane (note that the collection itself  fills a three
dimensional space).

In its ``ordered state", on which we focus here, the collection of moving entities has a non-zero average velocity in the thermodynamic ($N\rightarrow\infty$, where $N$ is number of self-propelled entities in the system) limit.

The ``sliding columnar phase" of cationic-lipid DNA complexes is a
conjectured phase in which nearly straight DNA molecules are confined between a set
of 3D space filling lipid layers. The DNA align with each other, both in a
given ``layer", and between layers. In addition, there are positional
interactions  between neighbors within a given layer, but only orientational
interactions between layers.

To establish the connection between the two  very different systems we've
just described, we first formulate a generic hydrodynamic theory of
easy-plane polar active fluids. Using a dynamical renormalization group
(DRG) analysis, we then show that this model can be mapped onto the
time-dependent-Ginsburg-Landau (TDGL) model of an equilibrium,
divergence-free 3D magnet with an easy-plane. We then focus on  the
static (equal-time) properties and use a static equilibrium RG analysis to
connect our magnet to an equilibrium system in the sliding columnar
phase. The connections between these theoretical models are
illustrated in Fig. \ref{fig:cartoon}. Through these connections we are able to go beyond  linear
hydrodynamics and calculate the exact scaling of the equal-time velocity correlation
function  of our active system, which is radically altered by non-linearities.

Specifically, we find  the equal-time velocity fluctuations in the active fluid are proportional  to the orientational fluctuations of the DNA strands in the sliding columnar phase. Since the latter are known to be bounded \cite{Column1,Column2,Column}, we can conclude that velocity fluctuations are bounded in the active fluid as well, implying that the ordered phase with non-zero average velocity exists. Furthermore, we can use the mapping to show that the
connected part of the velocity correlation function in the
ordered phase has the following equal-time scaling behavior at large distances
(i.e., $x\gg\xi_{x}$, $z\gg\xi_{z}$, or $y\gg\xi_y$):
%{\cred [JT: I think there should be three different crossover lengths $\xi_{x}$, $\xi_{y}$, and $\xi_{z}$ here, related by $\xi_{x}=\xi_{y}^2/\lambda_x$, and $\xi_{z}=\xi_{y}^2/\lambda_z$, where $\lambda_{x}\equiv\sqrt{K_{xx}\over B}$ and $\lambda_{x}\equiv\sqrt{K_{xz}\over B}$.  Do you guys agree? ]}{\cblue [LC: I agree.]}{\cmag [[CFL: Agreed.]]}
\begin{eqnarray}
&\langle\left[\bv(0,t)-\bv_0
\right]\cdot\left[\bv(\br,t)-\bv_0\right]\rangle\sim\nonumber\\
&\left\{
\begin{array}{ll}
{\left(\ln{X}\right)^{1\over 4}\ln\left(\ln X\right)\over \left(x\Lambda\right)^2}\,,&X\gg Y^{1\over 2}\left(\ln{Y}\right)^{5\over 16},Z\left(\ln{Z}\right)^{1\over 8}\\
{\left(\ln{Y}\right)^{-{3\over 8}}\over |y|\Lambda}\,,&Y\gg X^2\left(\ln{X}\right)^{-{5\over 8}},Z^2\left(\ln{Z}\right)^{-{3\over 8}}\\
{1\over \left(z\Lambda\right)^2}\,,&Z\gg X\left(\ln{X}\right)^{-{1\over 8}},Y^{1\over 2}\left(\ln{Y}\right)^{3\over 16}
\end{array},
\right.\label{RealCorr}
\end{eqnarray}
where ${\br} = (x,y,z)$, $\xi_{x,y,z}$ are some non-universal lengths which we calculate in \cite{SM}, $\Lambda$ is the ultra-violet cutoff, $X\equiv {|x|\over\xi_{x}}$, $Y\equiv {|y|\over\xi_y}$, $Z\equiv {|z|\over \xi_{z}}$, $\bv_0$ is the average velocity of the system taken to be along the $x$-direction, and the easy plane is denoted as the $xy$-plane. At short distances (i.e., $|x|\ll\xi_{x}$, $|y|\ll\xi_y$,  and $|z|\ll\xi_{z}$) (\ref{RealCorr}) reduces to that of linear theory, all the logarithms becoming 1. Note that the crossover lengths $\xi_{x}$, $\xi_{z}$ and $\xi_y$ can be very large, since they are exponential functions of the parameters - in particular, of the noise strength defined below. Therefore, to observe the logarithms in (1) in experiments or simulations, very large system sizes may be required. For smaller systems, the correlations behave as in (1), but with all of the logarithms replaced by constants.

%\sout{	Besides establishing the mapping between a non-equilibrium,
%active system and an equilibrium system, our work is also of relevance
%to the physics of a swarm of motile synthetic or biological agents that
%have a propensity to move on planes normal to a common axis. }
%
%\sout{
%Not all active particles move equally along any direction in the space; Some of them tend to move parallel to certain plane which we name as ``easy plane''; for instance, birds are more likely to move horizontally than vertically and in this case the  horizontal plane is the easy plane. On the other hand, recent studies show that incompressible polar active fluid exhibits interesting physics \cite{chen_njp15, chen_NC}.
%}
%
%\sout{
%In this paper, we
%formulate a hydrodynamic (i.e., long-wavelength and long-time) theory of the ordered,
%moving phase of a 3D incompressible easy-plane polar active fluid.
%We find that this theory also applies to an equilibrium problem:
%a divergence-free 3D easy-plane magnets. The Hamiltonian of the latter can be further  mapped onto that of the sliding columnar phase \cite{Column}.
%Through this connection we are able to go beyond the linear hydrodynamics and work out the equal-time velocity correlation function.
%}

%By definition (\ref{RealCo}) the real space fluctuations of the velocity are given by $C_r(\mathbf 0,t)$, which can be shown to be finite. This confirms that the system can develop long-range orientational order.
To formulate a hydrodynamic  theory of 3D incompressible easy-plane
polar active fluids, we start with the generic equation of motion (EOM) of  compressible active
fluids \cite{TT1,TT3,TT4, rean}, with an addition linear damping term that
forces the velocity field to lie preferentially on the easy plane ($xy$-plane):
\beqn
\partial_t\rho &=& -\nabla\cdot(\bv\rho) \ ,
\label{Reconservation}
\\
\nonumber
\partial_{t}\bv&=&-2g v_z\hat{{\bf z}}-\lambda_1(\bv\cdot\nabla)\bv-\lambda_2(\nabla\cdot\bv)\bv-\lambda_3 \nabla(v^2)
\\
\nonumber
&&+U\bv -\nabla P -\bv\left( \bv \cdot \nabla P_2 \right) +\mu_{{\rm B}} \nabla(\nabla\cdot \bv)\nonumber\\
	&&+ \mu_{\rm T}\nabla^{2}\bv +\mu_{2}(\bv\cdot\nabla)^{2}\bv +{\mathbf{f}}\, .
	\label{reEOM}
\eeqn
Here $\bv(\br,t)$, and $\rho(\br,t)$ are respectively the coarse grained
continuous velocity and density fields, and  $(-2g v_z\hat{{\bf z}})$ is
a symmetry breaking damping term that makes the velocity field tend to lie
parallel to the $xy$-plane. The  random driving force $\mathbf{f}$  is
assumed to be Gaussian with white noise correlations:
\begin{eqnarray}
\label{eq:noise}
\la f_m(\br,t)f_n(\br',t')\ra=2D%\Delta
\delta_{mn}\delta^{d}(\br-\br')\delta(t-t')
\label{white noise}
\end{eqnarray}
where the ``noise strength" $D$ is a constant parameter of the system,
and $m$, $n$
denote Cartesian components. Note that since our intention is to
study systems that are not momentum conserving, these chosen statistics
do not conserve momentum, in contrast to thermal fluids (e.g., Model A in
\cite{FNS}).

All of the parameters $\lambda_i\  (i = 1 \to 3)$,
$U$, the ``damping coefficients" $\mu_{\rm B,T,2}$, the  ``isotropic pressure'' $P(\rho,
v)$ and the  ``anisotropic pressure'' $P_2 (\rho, v)$
are, in general, functions of the density $\rho$ and the
magnitude $v\equiv|\bv|$ of the local velocity.

Since we are interested in the ordered state (which we will show to exist
later), we assume the $U$ term makes the local
$\bv$ have a nonzero magnitude $v_0$
in the steady state, by the simple expedient of having $U>0$ for $v<v_0$,
$U=0$ for $v=v_0$, and $U<0$ for $v>v_0$. We treat
fluctuations by expanding
$\bv$ around $v_0\hat{\bf x}$, thus defining  $\bu(\br, t)$ as the
small fluctuation in the velocity field about this mean:
\beqn
\bv(\br, t) =(v_0+u_x(\br, t))\hat{\bf x} + u_y(\br, t) \hat{\bf y}+ u_z(\br, t) \hat{\bf z}
\ .\label{Expansion1}
\eeqn

We now go to %take
the incompressible limit by taking  the isotropic pressure $P$ %{\it only}
to be extremely sensitive to departures from the mean density $\rho_0$,
so that it suppresses density fluctuations extremely effectively.
%\sout{One could alternatively consider making $U(\rho, v)$ and
%$P_2(\rho, v)$	extremely sensitive to changes in $\rho$ as well.
%We will discuss these possibilities in \cite{SM}.}
%{\cmag [[CFL: I suggest	saving this interesting discussion about how to go
%to the incompressible limit for another paper, maybe a PRE. My reason is
%not to milk another paper out of this, but rather, I think this interesting
%discussion deserves a proper appearance, instead of being buried in the
%SM. What do you think?]]}{\cred JT: I agree. }	
%\sout{Focusing here on the case in which {\it only} the isotropic pressure
%$P$ becomes extremely sensitive to changes in the density, we see that,
%in this limit, in which the isotropic pressure suppresses density fluctuations
%extremely effectively,}
Therefore, changes in the density are
too small to affect  $U(\rho, v)$, $\lambda_{1,2,3}(\rho, v)$, $\mu_{{\rm B},
{\rm T},2}(\rho, v)$, and $P_2(\rho, v)$.  As a result,
all of them effectively become functions only of the
speed $v$; their $\rho$-dependence drops out since $\rho$ is
essentially constant. Another consequence of the suppression of density
fluctuations by the isotropic pressure $P$ is that the continuity
equation (\ref{Reconservation}) reduces to the condition
\beq
\label{eq:incomp}
\nabla \cdot\bv=0
\ .
\eeq
In particular, the $\lambda_2$ and $\mu_B$ terms vanish due to this  condition (\ref{eq:incomp}).

Models of incompressible active fluids as defined here are rich in physics:
incompressible active fluids whose their motion is {\it not} confined to an
easy plane, undergo a critical order-disorder transition that exhibits novel
universal behavior \cite{us1}; and in the ordered phase in 2D, the system
can be mapped onto the (1+1)D Kardar-Parisi-Zhang surface growth
model \cite{us2}. With the easy-plane restriction considered here, we
shall see that the ordered phase can be mapped onto  an equilibrium
soft matter system.

%where $\perp$ denotes the $xy$ plane. This condition can, as in simple fluid %mechanics, be used to determine the isotropic pressure $P$.

With the incompressibility condition (\ref{eq:incomp}) taken into account  and
 using (\ref{Expansion1}) in (\ref{reEOM}), we find
%	\begin{eqnarray}
%	\pp_t u_m &=& f_m-\pp_m P -2\alpha u_x\delta_{mx} -2gu_z\delta_{mz}
%	- \lambda^0_1v_0\pp_x u_m
%	\nonumber
%	\\
%	&&
%	- \lambda^0_4v_0^3\delta_{xm}(\pp_x u_x)+\mu^0_{\rm T} \nabla^2 u_m
%	+  \mu_2^0v_0^2 \pp_x^2u_m\nonumber
%	\\
%	&&-{\alpha\over v_0}\left({u_y^3\over v_0}\delta_{my}+2u_xu_y\delta_{my}+u_y^2\delta_{mx}\right)
%	\nonumber
%	\\
%	&&
%	-\lambda^0_1u_y\pp_yu_{y}\delta_{my}
%	\ , \label{ReEOMNL1}
%	\end{eqnarray}
\beqn
\partial_{t}\bu&=&-2g u_z\hat{{\bf z}}-\lambda^0_1v_0\pp_x\bu-\lambda^0_1(\bu\cdot\nabla)\bu
\nonumber\\
&&-2\alpha \left(u_x+{u_y^2\over 2 v_0} \right)\hat{{\bf x}}
 -\nabla P
 \nonumber\\
	&& +\left[ \mu^0_{\rm T} \left(\pp_y^2+\pp_z^2\right)
	%+\mu \pp_x^2 	+	 \left[\mu_{\rm T}^{\cblue 0}{\cblue \left(\pp_y^{2}+	\pp_z^{2}\right)}
+\mu_{x}\pp_x^{2}\right]\bu
	+{\mathbf{f}}\, .
	\label{uEOM}
\eeqn
%		\begin{eqnarray}		
% \label{ReEOMNL1}
%	\pp_t u_x &=& f_x-\pp_x P -2\alpha u_x-\frac{\alpha}{v_0} \left(u_y^2+u_z^2\right)		
%	\\
%&&
%-\lambda^0_1v_0\pp_x u_x
%	- \lambda^0_4v_0^3\pp_x u_x+\mu^0_{\rm T} \nabla^2 u_x
%	+  \mu_2^0v_0^2 \pp_x^2u_x
%	 \ ,
%	\nonumber
%	\ , \label{ReEOMNL1}
 %	\end{eqnarray}
%		\begin{eqnarray}	
%\pp_t u_m &=& f_m-\pp_m P-2g u_z \delta_{mz} -\frac{2\alpha}{v_0} \left(u_x +\frac{u_y^2+u_z^2}{2v_0}\right)u_m
%	\nonumber
%\\
%&&
%-\lambda^0_1v_0\pp_x u_m -\lambda^0_1 \left( u_y\pp_y u_m+u_z\pp_z u_m \right)
%\nonumber
%\\
%&&
%+\mu^0_{\rm T} \nabla^2 u_m
%+  \mu_2^0v_0^2 \pp_x^2u_m
% \label{ReEOMNL1m}
%\end{eqnarray}
where we have defined
$\alpha\equiv{1\over2}{dU\over dv}\big|_{v=v_0}$,
$\mu_x\equiv \mu_{\rm T}^0+\mu^0_2 v_0^2$, and have absorbed a
term $W(v)$ into the pressure $P$, where $W(v)$ is derived from
$\lambda_3(v)$ by solving $ {1\over 2v}{dW\over dv}=\lambda_3(v)$.
The superscript ``0'' means that the $v$-dependent coefficients are evaluated
at $v = v_0$.
In the above EOM, we have also omitted ``obviously" irrelevant terms
%(such as one proportional to
%${dP_2(v)\over dv}\pp_xu_x\hat{{\bf x}}$, which is negligible relative to the
%$\alpha u_x\hat{{\bf x}}$ term that we've kept)
in the sense discussed in \cite{us2}.
%{\cred [LC: I suggest that we hide the discussion in the
%parenthesis. Because the reader can get confused: then why do we still
%keep $\lambda_1v_0\partial_xu$ in the equation. If we really want to
%explain this issue thoroughly, we can do it in the long paper. When I
%brought out this demon. I just wanted to make sure I understood what
%we did as best as I can. Sadly I often raise questions but am not good at
%solving them myself.}

We can further simplify the EOM by making a Galilean transformation  to a ``pseudo-co-moving" co-ordinate system moving in the direction $\hat{\bf x}$ of mean flock motion at speed %$v_1 \equiv
$\lambda^0_1v_0$
to eliminate the ``convective term"  $\lambda^0_1v_0\pp_x\bu$
from the right hand side of (\ref{uEOM}); this
leaves us with our final simplified form for the EOM:
\beqn
\partial_{t}\bu&=&-2g u_z\hat{{\bf z}}-\lambda^0_1(\bu\cdot\nabla)\bu -2\alpha \left(u_x+{u_y^2\over 2 v_0} \right)\hat{{\bf x}}
\nonumber\\
&&-\nabla P+
\left[ \mu^0_{\rm T} \left(\pp_y^2+\pp_z^2\right)
%\mu_{\rm T}^{\cblue 0}{\cblue \left(\pp_y^{2}+	\pp_z^{2}\right)}
+\mu_{x}\pp_x^{2}\right]\bu+{\mathbf{f}}\, .
	\label{uEOMfinal}
\eeqn

Since both $u_x$ and $u_z$ are ``massive", because of the $2gu_z$ and $2\alpha u_x$ terms, respectively,  we expect the fluctuations of $u_y$ to dominate over those of $u_{x,z}$. We therefore focus on the EOM of $u_y$. We obtain this  by Fourier transforming (\ref{uEOMfinal}) in space at wavevector $\bf{q}$ and eliminating the pressure term
%in (\ref{ReEOMNL1})
by acting on both sides of the equations with the
transverse projection operator $P_{lm}(\bq)=\delta_{lm}-q_lq_m/q^2$, and looking at the $l=y$ component of the resulting equation.
The linearized EOM of $u_y$ thereby becomes
%{\cred [LC: I have added the non-linear terms since they will be mentioned later.]}
%\begin{widetext}
%	\begin{eqnarray}
%	\partial_t u_y(\bq,t)&=& -iv_0\lambda^0_1q_xu_y(\bq,t)+i\lambda^0_4v_0^3{q_x^2q_y\over q^2}u_x(\bq,t)-\Gamma(\bq)u_y(\bq,t)+P_{yx}(\bq)\mathcal{F}_{\bq}\left[-2\alpha
%	\left(u_x(\br,t)+\frac{u_y^2(\br,t)+{\cmag u_z^2(\br,t)}}{2v_0}\right)%- \lambda^0_4v_0^3(\pp_x u_x(\br,t))
%	\right]\nonumber\\
%	&&+2g{q_yq_z\over q^2}u_z(\bq,t)+P_{yy}(\bq)\mathcal{F}_{\bq}\left[-{\alpha\over v_0}\left({u_y^3\over v_0}+2u_xu_y\right)-\lambda_1^0u_y\pp_yu_y\right]+P_{ym}(\bq)f_m(\bq,t)
%	\ ,
%	\label{ReFullEOM1}
%	\end{eqnarray}
%\end{widetext}
	\begin{eqnarray}
	\partial_t u_y(\bq,t)&=&P_{ym}(\bq)f_m(\bq,t)-\Gamma(\bq) u_y(\bq,t)
	\nonumber
	\\
&&+2\alpha {q_xq_y\over q^2}u_x(\bq,t)+2g{q_yq_z\over q^2}u_z(\bq,t)
%\nonumber
%	\\	
%	&&	
%-\lambda^0_1 \left( u_y\pp_y u_y+u_z\pp_z u_y \right)
%\nonumber
%	\\	
%	&&	
%-\frac{2\alpha}{v_0} \left(u_x +\frac{u_y^2}{2v_0}\right)u_y
%		\nonumber
%	 +\ii v_0\lambda^0_1q_x u_y(\bq,t)
%	 -\ii\lambda^0_4v_0^3{q_x^2q_y\over q^2}u_x(\bq,t)
	 \ ,
	\label{linEOMy}
	\end{eqnarray}
where
%where $\mathcal{F}_{\bq}$ represents the
%Fourier component at wavevector $\bq$, i.e.,   $\mathcal{F}_{\bq}[g(\br)] \equiv \int \dd^3 r\, g(\br) e^{-\ii \bq \cdot \br}$; and
%\begin{eqnarray}
we've defined $\Gamma(\bq)\equiv\mu_x q_x^2+\mu_{\rm T}^0\left(q_y^2+q_z^2\right)$. %\, ,\label{Gamma1SM}
%\end{eqnarray}
%with $\mu\equiv \mu_{\rm T}^0+\mu_2^0v_0^2$.

To proceed further, we perform a DRG
analysis by first  rescaling  the coordinates and fluctuating fields:
\beqn
x &\mapsto & \ee^{\ell} x \,,~
y \mapsto  \ee^{\zeta_y\ell} y \,,~
z\mapsto \ee^{\zeta_z\ell}z\,,~
t \mapsto  \ee^{w\ell} t\,,
\label{Re2dResclx}
%\\
%&&u_x\mapsto   e^{\chi_x\ell} u_y \,,~
%u_y \mapsto  e^{\chi_y\ell} u_x \,,~
%%\label{Re2dResclux}
%u_z\mapsto  e^{\chi_z\ell} u_z
%\ ,
\\
u_y &\mapsto & \ee^{\chi_y\ell} u_y \,,\\
\label{Re2dResclux}
u_x &\mapsto & \ee^{\chi_x\ell} u_x=\ee^{\left(\chi_y+1-\zeta_y\right)\ell} u_x\,,\\
u_z &\mapsto & \ee^{\chi_z\ell} u_z =\ee^{\left(\chi_y+\zeta_z-\zeta_y\right)\ell} u_z \, ,
\label{Re2dRescluz}
\eeqn
where we've enforced the equalities in (\ref{Re2dResclux}) and
(\ref{Re2dRescluz})
%\sout{ follows from comparing $\pp_x u_x$ with $\pp_y u_y$ based on}
to maintain the form of the incompressibility
condition (\ref{eq:incomp}).

Applying these rescalings (\ref{Re2dResclx}-\ref{Re2dRescluz}) to
the linear EOM of $u_y$ (\ref{linEOMy}),
we find that
%\sout{the linearized EOM remains invariant  if we re-scale
%the coefficients accordingly as shown below}
the parameters rescale as
 \begin{eqnarray}
&& \mu_x \mapsto  \ee^{\left(w-2\right)\ell} \mu_x \ ,
\sep
\mu_{\rm T}^0\mapsto  \ee^{\left(w-2\zeta_z\right)\ell} \mu_{\rm T}^0 \ ,
\\
&& \alpha \mapsto  \ee^{\left(w-2\zeta_y+2\min(1 ,\zeta_y, \zeta_z)\right)\ell} \alpha \ ,
\\
&&
g \mapsto  \ee^{\left(w-2\zeta_y+2 \min(1 , \zeta_y, \zeta_z)\right)\ell} g \ , \\
&& D \mapsto  \ee^{\left(w-2\chi_y-\zeta_y-\zeta_z-1\right)\ell} D\, .
\end{eqnarray}
%\sout{In the above power counting exercise, we have used the fact that
%$\chi <0$ so that the ordered phase exists, and the reasoning that since
%fluctuations in $u_y$ dominates, we expect $|\chi_y|<|\chi_{x,y}|$, which
%implies  $\zeta_y>1$ and $\zeta_y>\zeta_{z}$ via \eqs (\ref{Re2dResclux}) \& (\ref{Re2dRescluz}). As a result,}
where the
 ${\rm min}(1, \zeta_y, \zeta_z)$ appears because,  in the limit $\ell\to\infty$, the values of
 $\zeta_{y,z}$ determine %\sout{in the limit $\ell\to\infty$}
 which
 component $q_{x,y,z}$ dominates the $q^2$'s that appear in (\ref{linEOMy}).

%We now use the standard {\cmag DRG analysis} % renormalization group logic
%to assess the importance of the non-linear terms in (\ref{ReEOMNL1}) \& (\ref{ReEOMNL1m}).
%The logic  is
We now choose   the rescaling exponents $w$, $\zeta_{y,z}$, and $\chi_y$ so as to keep the size of the fluctuations in the  field $\bu$ fixed upon rescaling. This is %clearly
accomplished by keeping  $\alpha$, $g$,  $\mu_x$, $\mu_{\rm T}^0$, and $D$ fixed. From the rescalings just found, this leads to four simple linear
equations in the four unknown exponents $w$, $\zeta_{y,z}$, and $\chi_y$; solving these, we find
%the values of these exponents in the linearized theory:
\begin{eqnarray}
w=\zeta_y=2\,,~\zeta_z=1\,,~\chi_{_{y}}=-1\,.\label{LinearExp}
\end{eqnarray}
With these exponents in hand, we can now assess the importance of the
non-linear terms in the full EOM for $u_y$ at long length scales, simply by looking at how their coefficients rescale.
%(We don't have to worry about the size of the actual non-linear terms themselves changing upon rescaling, because we have chosen the rescalings to keep them constant in the linear theory.)
We find that all the non-linear terms whose coefficients are proportional to $\alpha$ are ``marginal", that is, the coefficients of these terms remain fixed upon rescaling. However, the last remaining non-linear term is ``irrelevant''  because its coefficient gets smaller upon rescaling: $\lambda^0_1 \mapsto  \ee^{-\ell} \lambda^0_1$.
%\begin{eqnarray}
%\lambda^0_1 \mapsto  \ee^{-\ell} \lambda^0_1\,.
%\end{eqnarray}
Hence, this term will not affect the long-distance behavior, and can be dropped from the problem.

Now we come back to Eq. (\ref{uEOMfinal}) and drop the
``irrelevant'' $\lambda_1^0$ non-linear term.
The reduced EOM becomes identical to the TDGL model:
\beq
\frac{\pp u_m}{\pp  t} =-\frac{\delta H}{\delta u_m}+f_m
\label{eq:TDGL}
\ ,
\eeq
 where $\mathbf f$ is the thermal noise whose statistics are described by Eq.\ (\ref{white noise}) with $D=k_BT$, %=1/\beta$,
 and  with the Hamiltonian
%\beq
%H'=H- \int \dd^3r \ P(\br)(\mathbf {\nabla} \cdot \mathbf u)
%\label{Hconst}
%\ .
%\eeq
%In Eq. (\ref{Hconst}) {\cblue $H$ is the Hamiltonian for an ``easy-plane" magnet:
%\beqn
%H&=&\frac{1}{2} \int \dd^3r \Bigg\{V(|\mathbf M(\bu)|)+
%\mu_{\rm T}^0|\vec{\nabla} \mathbf M(\bu)|^2%\mu|\nabla u_y|^2
%\\
%&&
%+%\left(\mu_x-\mu_T^0\right)
%{\cmag \mu_x}|\pp_x \mathbf M(\bu)|^2+gM(\bu)_z^2\Bigg\}\, ,
%\label{ReHxy1}
%\eeqn
%}where we've defined $\mathbf M\equiv\left(v_0+u_x\right)\hat{x}+u_y\hat{y}+u_z\hat{z}$, and $V(|\mathbf M|)$ is minimized at $|\mathbf M|=v_0$.
%Due to the presence of the anisotropic mass term $gM_z^2$ this Hamiltonian has a overall minimum at $|\mathbf M_{\perp}|=M_0$, $M_z=0$.  Expanding about this minimum to leading order in $\bu$,
%and keeping only ``relevant" terms, we get
\beqn
H&=&\frac{1}{2} \int \dd^3r \Bigg\{2\alpha\left(u_x+{u_y^2\over 2v_0}\right)^2+ 2gu_z^2%\mu|\nabla u_y|^2
\nonumber\\
&&+%{\cblue\left(\mu_x-\mu_T^0\right) }
\left(\mu_x-\mu_{\rm T}^0\right)(\pp_x u_y)^2+ \mu_{\rm T}^0 |\nabla u_y|^2
\Bigg\}\, ,
\label{ReHxy2}
\eeqn
where
%\sout{we've defined the ``longitudinal mass''
%$2\alpha\equiv \left.\left(\partial^2 V\over\partial |\bM|^2\right)\right|_{|\bM|=v_0}$. We have dropped the anharmonic terms which are proportional to either $u_x^2$ or $u_z^2$ since they are ``irrelevant" compared to either $u_x^2$ or $u_z^2$ in the harmonic part of the Hamiltonian.
%In addition to $H$ in \eq (\ref{Hconst}),}
we have
$P(\br)$ as the Lagrange multiplier employed to enforce the
divergence-free (incompressibility) constraint $\nabla \cdot \bu=0$.
One can now straightforwardly check that the TDGL model equation in
(\ref{eq:TDGL}) does lead to the EOM (\ref{uEOMfinal}) without the
``irrelevant" $\lambda_1^0$ non-linear term.
	
This mapping between a nonequilibrium active fluid model and
a ``divergence-free'' easy-plane equilibrium magnet model allows us to investigate the fluctuations in our original active fluid model by  studying the partition function of the equilibrium model (\ref{ReHxy2}).
%{\cred \sout{Focusing therefore on the partition function, we first note that in the Hamiltonian $H$ (\ref{ReHxy2}), $u_y$ is coupled to $u_x$ via the coefficient $\alpha$ but not to $u_z$. Since we expect fluctuations in $u_y$ to dominate,}
Since the magnetization prefers to point parallel to
the $xy$ plane, the fluctuations of $u_z$ are expected to be much smaller than those of $u_{x,y}$. %\sout{make here the ansatz}
% {\cred [[JT: The term "ansatz" is usually reserved for a specific mathematical form that's inserted as a trial into some mathematical expression. I therefore prefer to use the phrase "we here assume".]]
We here assume that the fluctuations in $u_z$  become negligible upon coarse-graining, i.e., $\lim_{\ell \rightarrow \infty} g=\infty$.
We will justify this assumption self-consistently later.

With the elimination of $u_z$ (due to the divergence of $g$ upon RG transformation), we can now introduce the streaming function $h$ to enforce the incompressibility condition:
%To deal with the exact identity $\mathbf{\nabla}_{\perp} \cdot \mathbf u=0$, we can use a trick familiar from the study of incompressible fluid mechanics: we introduce a ``streaming function"; i.e., a new scalar field $h(\br)$ such that
\beq
u_x = -v_0 \pp_y h
\sep
u_y = v_0 \pp_x h
\ .
\label{VTrans1}
\eeq
Since this construction guarantees that the incompressibility condition $\mathbf \nabla \cdot \mathbf u=\pp_x u_x+\pp_y u_y=0$ is automatically satisfied (once we set $u_z=0$), there is no constraint on the field $h(\br)$.

%Making  the  substitution (\ref{VTrans1}), the Hamiltonian  (\ref{ReHxy2})
%(\ref{Hxy2})
Substituting  (\ref{VTrans1}) into the Hamiltonian  (\ref{ReHxy2}), it becomes
%becomes
(ignoring irrelevant terms like $(\pp_x\pp_y h)^2$, which is irrelevant compared to $(\partial_y h)^2$ because it involves two extra $x$-derivatives)
\begin{eqnarray}
%H_s={1\over 2}\int d^3r \left[B'\left(\partial_y h-{(\partial_x h)^2\over 2}\right)^2\right.\nonumber\\
%\left.+K'_{xx}(\partial^2_x h)^2
%+K'_{xy}(\partial_x\partial_z h)^2\right]
%\ ,\label{Sli-columnar}
H_s={1\over 2}\int d^3r \left[B\left(\partial_y h-{(\partial_x h)^2\over 2}\right)^2\right.\nonumber\\
\left.+K_{xx}(\partial^2_x h)^2
+K_{xz}(\partial_x\partial_z h)^2\right]
\ ,\label{Sli-columnar}
\end{eqnarray}
where $B=2\alpha v_0^2$, $K_{xx}=\mu_x$, and
$K_{xz}=\mu_{\rm T}^0$.
This Hamiltonian is exactly the elasticity theory for the sliding
columnar phase \cite{Column1,Column2,Column}, with columns oriented along $x$, sandwiched
between rigid plates which stack along $z$, fluctuating with
displacements $h$ restricted along $y$ \cite{Column} (see \fig \ref{fig:cartoon}).
%\sout{{\cmag We have thus added the subscript $s$ to indicate this fact.}}
In the sliding columnar phase, there are
%\sout{strong correlations}
interactions between the orientations of the columns in different layers but none between the positions. The elastic coefficients $B$, $K_{xx}$, and $K_{xz}$ are respectively the compression, bend, and twist moduli \cite{footnote1}.

It has been shown that the anharmonic terms in (\ref{Sli-columnar}) lead to an infinite renormalization of the elastic constants \cite{Column},
as also happens in  the 3D smectic phase \cite{GP}. Specifically, in the long wavelength
limit (i.e.,  $|q_{x,y,z}|< \xi_{x,y,z}$,
%$\sqrt{q_x^2+q_z^2}\ll \xi_{xz}^{-1}$ %\xi_{xz}^{(h)}$
%and $|q_y|\ll \xi_y^{-1}$ %\xi_y^{(h)}$
where $\xi_{x,y,z}$ %($\xi_y=\Lambda\xi_{xz}^2$)
are  non-universal lengths %$\sout{\xi_{xz}^{(h)}}
which we  estimate in \cite{SM}),  the compression modulus vanishes logarithmically and  the bend
and twist moduli diverge logarithmically according to
%\begin{eqnarray}
%K'_{xx}(\bq)&\sim& \left[\ln(\Lambda/q)\right]^{{1\over 4}}\,,\label{ReK_xx}\\
%K'_{xz}(\bq)&\sim& \left[\ln(\Lambda/q)\right]^{{1\over 2}}\,,\label{ReK_xz}\\
%B'(\bq)&\sim& \left[\ln(\Lambda/q)\right]^{-\frac{3}{ 4}}\,,
%\label{ReB}
%\end{eqnarray}
\beq
K_{xz}(\bq)\sim K_{xx}^{\frac{1}{2}}(\bq) \sim B^{-\frac{1}{3}}(\bq) \sim |\ln q|^{\frac{1}{ 4}}
\, .
\label{ReB}
\eeq
In terms of the
coefficients of the active fluid model
%in (\ref{ReEOMNL3}) \& (\ref{ReEOMNL3m}),
we have
	\beq
	\mu_{\rm T}^0(\bq)\sim \mu_{x}^{\frac{1}{2}}(\bq) \sim \alpha^{-\frac{1}{3}}(\bq) \sim |\ln q|^{\frac{1}{ 4}}\, .
	\label{coeff_scal}
	\eeq	

%{\cmag As a result,}
%%Subsequently
%the equal-time
%correlation function of $h$ is given by
%\begin{eqnarray}
%\left<h(\bq,t)h(\bq',t)\right> =
%{\left(2\pi\right)^3D\delta\left(\bq+\bq'\right)\over B(q) q_y^2+K_{xx}(q) q_x^4+K_{x{\cmag y}}(q)q_{\cmag y}^2q_x^2}\,.\nonumber\\
%\end{eqnarray}
%

%Consequently,
%we should use renormalized coefficients in the expressions for the $\bu$-$\bu$ correlation function. Adding the three components (i.e., Eqs. (\ref{Reuxprop},\ref{Reuyprop},\ref{Reuzprop})) together

Using these renormalized parameters in our effective equilibrium model (\ref{ReHxy2}) we can obtain the full equal-time correlation function  for our original problem \cite{SM}:
\begin{eqnarray}
&&\langle\bu(\bq)\cdot\bu(\bq')\rangle\nonumber\\
&&={\left(2\pi\right)^3D\left[2\alpha(\bq)q_z^2+g\left(q_x^2+q_y^2\right)
\right]\delta(\bq+\bq')\over \left[\mu_x(\bq)q_x^2+\mu_{\rm T}^0(\bq)q_z^2\right] \left[2\alpha(\bq)q_z^2+gq_x^2\right]+2g\alpha(\bq)q_y^2}~,\nonumber \\
\label{CoUU1}
\end{eqnarray}
where we have neglected terms which have higher powers in $q$ than
the present ones in the numerator.
%{\cmag [[CFL: Maybe we should lose the $gq_y^2$ term in the numerator
%since we have omitted terms of the same order.]]}{\cred JT: Let's keep it,
%because it enables us to use this form even for lenght scales less than
%the non-linear lengths.}
Note that given the form of the above correlation function, we can now also conclude that the real space fluctuations
 	$\la u^2(\br)\ra \propto \int \dd^3 \bq \la| \bu(\mathbf q)|\ra$ is finite, thus implying the existence of long-ranged order in the divergence-free easy-plane magnet, as well as in our incompressible active fluid model.

%
%{\cmag Although the coefficients in the correlation function (\ref{CoUU1}) vary with $q$ in the small $q$ limit (\ref{ReB}), these coefficients approach a constant at shorter length scales (large $q$ limit).}
%%However, at shorter length scales the harmonic theory still holds.
%The crossover lengths which separate the two distinct behaviors can be worked out in the following way. We perform the ordinary perturbation theory and calculate the graphical corrections to the coefficients of the harmonic terms. Then we set the infrared cutoff to be the inverse of the crossover length such that the graphical correction is comparable to the coefficient itself.
%% For instance, we equate $\delta \alpha$ given by (\ref{ReDelta_alpha}) to $\alpha$ and solve the resultant equation for $L$, which gives the crossover length along $x$ and $z$:
%{\cmag By doing so, we find \cite{SM}}
%\begin{eqnarray}
%\xi^{(s)}_{xz}&=&\Lambda^{-1}\exp\left[16\sqrt{2}\pi^2v_0^2\mu^{3\over 2}\over D\alpha^{1\over 2}\Gamma\left(2\alpha\over g\right)\right]
%\\
%%\end{eqnarray}
%%Alternatively, when doing the integral in (\ref{ReDelta_alpha}) we place the infrared cutoff on $q_y$ instead of $q_{x,z}$ and then apply the same procedure. This gives the crossover length along $z$:
%%\begin{eqnarray}
%\xi^{(s)}_y&=&\Lambda\left[\xi^{(s)}_{xz}\right]^2=\Lambda^{-1}\exp\left[32\sqrt{2}\pi^2v_0^2\mu^{3\over 2}\over D\alpha^{1\over 2}\Gamma\left(2\alpha\over g\right)\right]\,.
%\end{eqnarray}
%

Given the logarithmic corrections in (\ref{ReB}), we will now justify the divergence of $g$ in the Hamiltonian (\ref{ReHxy2}) upon RG transformation, and thus the neglect of $u_z$. Specifically, in
%To this end, we now perform a RG analysis on the partition function.
 terms of the re-scalings in \eqs (\ref{Re2dResclx})--(\ref{Re2dRescluz}), the RG flow equations are:
 %{\cred I expressed $\chi_{x,z}$ in terms of $\chi_y$ and $\zeta_y,z$ in the first two flow equations} {\cblue JT agrees that's a good idea, but thinks we need to explain that we've done that; otherwise, these equations seem to appear by magic. I've attempted such an explanation after the equations  (in {\cred red}); let me know what you think.}
\begin{eqnarray}
{\dd\ln\alpha\over \dd\ell}&=&2\chi_y+3-\zeta_y+\zeta_z-\eta_{\alpha}\,,
\label{alpharr}
\\
%&=&2\chi_y+3+\zeta_z-\zeta_y-\eta_{\alpha}\,,\\
{\dd\ln g\over \dd\ell}&=&2\chi_y+1-\zeta_y+3\zeta_z\,,\label{Reg}\\
%&=&2\chi_y+1+3\zeta_z-\zeta_y\,,\\
{\dd\ln\mu_x\over \dd\ell}&=&2\chi_y-1+\zeta_z+\zeta_y+\eta_x\,,\\
{\dd\ln\mu_{\rm T}^0\over \dd\ell}&=&2\chi_y+1-\zeta_z+\zeta_y+\eta_z\,,
\end{eqnarray}
where $\eta_{\alpha,x,z}$ denote graphical corrections due to the anharmonic terms in the Hamiltonian (\ref{ReHxy2}),  and we've used the relation betweens between $\chi_y$ and $\chi_{x,z}$ implied by  (\ref{Re2dResclux}) and
(\ref{Re2dRescluz}) in (\ref{alpharr}) and (\ref{Reg}).
Note that there is {\it no} graphical correction to $g$.

We choose $\chi_y$ and $\zeta_{y,z}$ such that $\alpha$, $\mu_{x}$, and $\mu_{\rm T}^0$ are kept fixed. This choice fixes their values:
\beq
\chi_y=-1+{\eta_{\alpha}-\eta_z\over 4} \ , \
\zeta_z=1+{{\eta_z-\eta_x}\over 2}\ , \
\zeta_y=2-{\eta_x+\eta_{\alpha}\over 2}\ .\label{ScaExp}
\eeq
Plugging these scaling exponents into (\ref{Reg}) we get
\begin{eqnarray}
{\dd\ln g\over \dd\ell}=\eta_{\alpha}+\eta_z-\eta_x \label{RGflow_g}
\ .
\end{eqnarray}
From the logarithmic corrections in (\ref{ReB}), we deduce the
asymptotic behavior of $\eta_{\alpha,x,z}$ at large $\ell$:
\begin{eqnarray}
\eta_{\alpha}={3\over 4\ell}\,,~
\eta_x={1\over2\ell}\,,~
\eta_z={1\over 4\ell}\,.\label{etas}
\end{eqnarray}
Plugging these results into (\ref{RGflow_g}) we obtain $g\propto\sqrt{\ell}$,
which is consistent with the assumption we made earlier that $g$ flows to $\infty$.

Since we have justified that $u_z$ is negligible,
the flow lines of the incompressible ``easy-plane" polar active fluid are effectively restricted
to be parallel to the $xy$ plane. So the streaming function $h(\br)$ can be
viewed as the displacement field of the flow lines from their uniformly distributed
position in the steady state $\bv(\br)=v_0\hat{{\bf x}}$. Likewise for the magnets $h(\br)$
is the displacement field of the magnetic lines of flux.
Therefore, the mathematical connection between the theoretical models of the
incompressible ``easy-plane" polar active fluid,  ``easy-plane" magnets, and the sliding
columnar phase can be interpreted figuratively: the fluctuations of the flow lines, magnetic lines,
and the columns share the same scaling behavior at large length scale, as illustrated
in \fig \ref{fig:cartoon}.
%In the steady uniform state
%(i.e., $\bv(\br)=v_0\hat{x}$) the flow lines of the incompressible flock distribute
%uniformly over the space.

Using the RG transformation we can also work out the scaling behavior of the equal-time correlation
function $C(\br)\equiv\langle\bu(0,t)\cdot\bu(\br,t)\rangle$. The $C(\br)$ of the original system and the one of the rescaled system are connected by
%\begin{widetext}
\begin{eqnarray}
 &&C(\br)\nonumber\\
 &=&e^{2\int_0^{\ell_0}\chi_y d\ell}
        C\left(xe^{-\int_0^{\ell_0}d\ell},
        ye^{-\int_0^{\ell_0}\zeta_yd\ell},
        ze^{-\int_0^{\ell_0}\zeta_zd\ell}\right)\nonumber\\
        \label{Connect}
\end{eqnarray}
where the prefactor comes from the rescaling of $\bu$, which is
dominated by that of $u_y$. The exponents $\chi_y$ and $\zeta_{y,z}$ are
$\ell$-dependent and given by Eq. (\ref{ScaExp}). Note that
$\eta_{\alpha,x,z}$ are taken to be 0 for $\ell<\ell_n$ and given by
(\ref{etas}) only for $\ell>\ell_n$, where $\ell_n$ is determined by
the non-linear crossover lengths $\xi_{x,y,z}$.%$\xi_{xz,y}$.

For simplicity we first consider the special cases. For instance, for $x\neq 0$, $y=0$,
$z=0$ we choose $\ell_0=\ln\left(|x|\Lambda\right)$ and $\ell_n=\ln\left(  \xi_{x}\Lambda\right)$
and plug them into (\ref{Connect}). We find
\begin{eqnarray}
 C(x,0,0)\sim \left(x\Lambda\right)^{-2}\left[\ln\left(|x|\over  \xi_{x}\right)\right]^{1\over 4}\,.\label{Xscal}
        \label{}
\end{eqnarray}
Likewise we can obtain $C(\br)$ for the other two special cases, namely
$y\neq 0$, $x=0$, $z=0$ and $z\neq 0$, $x=0$, $y=0$:
\begin{eqnarray}
 &&C(0,y,0)\sim \left(y\Lambda\right)^{-1}\left[\ln\left(|y|\over\xi_y\right)\right]^{-{3\over 8}}\,,\label{Yscal}\\
 &&C(0,0,z)\sim \left(z\Lambda\right)^{-2}\,.\label{Zscal}
\end{eqnarray}
The crossover between these special cases can be obtained by equating
(\ref{Xscal},\ref{Yscal},\ref{Zscal}) to each other.

Alternatively, $C(\br)$ can be calculated more rigorously through
\begin{eqnarray}
C(\br)
=\int {d^3q\over \left(2\pi\right)^3}\int {d^3q'\over \left(2\pi\right)^3}
 \ \ \langle \bu(\bq,t)\bu(\bq',t)\rangle e^{i\bq\cdot\br}\,.
\end{eqnarray}
The details of this calculation are given in \cite{SM}.
The result from this approach agrees very well with that of the scaling argument
except that $C(x,0,0)$ acquires an extra multiplicative prefactor of $\ln\left(\ln x\right)$, which is such an extremely weak function of $x$ that it is unlikely to be detectable experimentally.
The scaling behavior of $C(\br)$ for arbitrary $\br$ is summarized in (\ref{RealCorr}).

\vspace{.1in}
%\section{Summary}
In summary, we have formulated a hydrodynamic theory of 3D
incompressible easy-plane polar active fluids. Using a DRG analysis we show that our active system in the ordered phase  is in the same universality class as the TDGL model of a modified type of easy-plane magnet in 3D. We then focus on the static (equal-time) properties of the system, and mapped it further onto an equilibrium system in the sliding columnar phase. Through these connections we were able to
 work out the singular wave vector dependence of the renormalized damping coefficients and the equal-time velocity correlation functions of our original model.

 Our work demonstrates  that for universal behavior, the boundary
 separating non-equilibrium and equilibrium systems can sometimes
 be blurry. We hope this will motivate further work  on identifying the key
 elements that distinguish nonequilibrium universality classes from
 equilibrium ones,  e.g., through investigating the signature of broken detailed balance \cite{battle} and the  amount of entropy production \cite{cates}.

 %Speaking of references, if we're going to submit this to PRL, the format of the references has to be changed. Unless they've changed their format, it's author (initials followed by last name with no comma), journal, volume, page, and year. Can I pass the buck of making that change to you guys?

LC acknowledges support by the National Science Foundation of China (under Grant No. 11474354);
JT thanks the Max Planck Institute for the Physics of Complex Systems in Dresden, Germany; the Department of Bioengineering at Imperial College, London; The Higgs Centre for Theoretical Physics at the University of Edinburgh, and the Lorentz Center of Leiden University, for their hospitality while this work was underway.

\onecolumngrid

\newpage
\begin{center}
	
	\textbf{\large Supplemental Materials:\\
		Squeezed in three dimensions, moving in two: Hydrodynamic theory of 3D
		incompressible easy-plane polar active fluids}
	\\
	\vspace{.1in}
	Leiming Chen\\
	{\it College of Science, China University of Mining and Technology, Xuzhou Jiangsu, 221116, P. R. China}\\
	Chiu Fan Lee\\
	{\it Department of Bioengineering, Imperial College London, South Kensington Campus, London SW7 2AZ, U.K.}\\
	John Toner\\
	{\it Department of Physics and Institute of Theoretical
		Science, University of Oregon, Eugene, OR $97403$}

\end{center}
\section{ calculation of the momentum-space equal-time correlation functions}
In the main text, we demonstrate how, when focusing on the equal-time properties of our active fluid model, one can map the system in the ordered phase onto the  equilibrium system of an easy-plane, divergence-free 3D magnet. In terms of the fluctuating flow fields $\bu$, the Hamiltonian is
\beq
\label{Seq:H}
H=\frac{1}{2} \int \dd^3r \Bigg\{2\alpha\left(u_x+{u_y^2\over 2v_0}\right)^2+gu_z^2%\mu|\nabla u_y|^2
+(\mu_x-\mu_{\rm T}^0) (\pp_x u_y)^2+ \mu_{\rm T}^0 |\nabla u_y|^2
\Bigg\}\, .
\eeq
We now write the Hamiltonian in Fourier space and use the incompressibility condition to eliminate $u_z$ in terms of $u_x$ and $u_y$:
\beq
u_z = -\frac{q_x}{q_z}u_x -\frac{q_y}{q_z}u_y
\ .
\eeq
This elimination also gets rid of the divergence-free constraint. Substituting the above relation into the Hamilton, and keeping now only quadratic terms, we have
\begin{eqnarray}
H&=&\frac{1}{2} \int {d^3q\over (2\pi)^3}\Bigg\{2\alpha u_x(\bq)u_x(-\bq) +g \left(\frac{q_x^2}{q_z^2} u_x(\bq)u_x(-\bq) -2\frac{q_x q_y}{q_z^2} u_x(\bq)u_y(-\bq) +\frac{q_y^2}{q_z^2} u_y(\bq)u_y(-\bq)\right)
+\mu_xq_x^2 u_y^2\nonumber\\
&&+ \mu_{\rm T}^0 \left(q_y^2+q_z^2\right) u_y^2
\Bigg\}\, .
\end{eqnarray}
%where $V$ is the volume of the system.
Within this harmonic approximation, the correlation of $u_{x,y}$ can be obtained by inverting the quadratic form in the integrand of the above Hamiltonian:
\beqn
\la u_x (\bq) u_x (\bq') \ra &=&{\left(2\pi\right)^3D\left[ gq_y^2+q_z^2 (\mu_x q_x^2+\mu_{\rm T}^0 q_z^2)
	\right]\delta(\bq+\bq')\over \left(\mu_xq_x^2+\mu_{\rm T}^0q_z^2\right) \left(2\alpha q_z^2+gq_x^2\right)+2g\alpha q_y^2}\,
\\
\la u_y (\bq) u_y (\bq') \ra &=&{\left(2\pi\right)^3D\left( 2\alpha q_z^2 +gq_x^2
	\right)\delta(\bq+\bq')\over \left(\mu_xq_x^2+\mu_{\rm T}^0q_z^2\right) \left(2\alpha q_z^2+gq_x^2\right)+2g\alpha q_y^2}\,
\\
\la u_x (\bq) u_y (\bq') \ra &=&{\left(2\pi\right)^3gDq_xq_y\,
	\delta(\bq+\bq')\over \left(\mu_xq_x^2+\mu_{\rm T}^0q_z^2\right) \left(2\alpha q_z^2+gq_x^2\right)+2g\alpha q_y^2}\,.
\eeqn
The correlation of $u_z$ can be calculated combining the above results with the incompressibility condition:
\beqn
\la u_z (\bq) u_z (\bq') \ra &=& \frac{q_x^2}{q_z^2} \la u_x (\bq) u_x (\bq') \ra -
2 \frac{q_xq_y}{q_z^2} \la u_x (\bq) u_y (\bq') \ra+ \frac{q_y^2}{q_z^2} \la u_y (\bq) u_y (\bq') \ra
\\
&=&{\left(2\pi\right)^3D\left[ 2\alpha q_y^2 +q_x^2 (\mu_x q_x^2+\mu_{\rm T}^0 q_z^2)
	\right]\delta(\bq+\bq')\over \left(\mu_xq_x^2+\mu_{\rm T}^0q_z^2\right) \left(2\alpha q_z^2+gq_x^2\right)+2g\alpha q_y^2}
\ .
\eeqn
The full correlation function of $\bu$ is therefore
\beqn
\la \bu (\bq)\cdot \bu(\bq') \ra&=&\la u_x (\bq) u_x (\bq') \ra+\la u_y (\bq) u_y (\bq') \ra+\la u_z (\bq) u_z (\bq') \ra
\\
&=&
{\left(2\pi\right)^3D\left[2\alpha q_z^2+g\left(q_x^2+q_y^2\right)\right]\delta(\bq+\bq')\over
	\left(\mu_x q_x^2+\mu_{\rm T}^0 q_z^2\right) \left(2\alpha q_z^2+gq_x^2\right)+2g\alpha q_y^2}
\ .
\label{uuSI1}
\eeqn
The effect of the anharmonic terms in the Hamiltonian or, equivalently, the nonlinear terms in the EOM)  can be incorporated by replacing the coefficients $\alpha$, $\mu_{\rm T}^0$, and $\mu_x$  with the  $q$-dependent  quantities given by equation (25) of  the main text.

\section{Estimation of the nonlinear crossover lengths}

\begin{figure}
	\begin{center}
		\includegraphics[scale=.32]{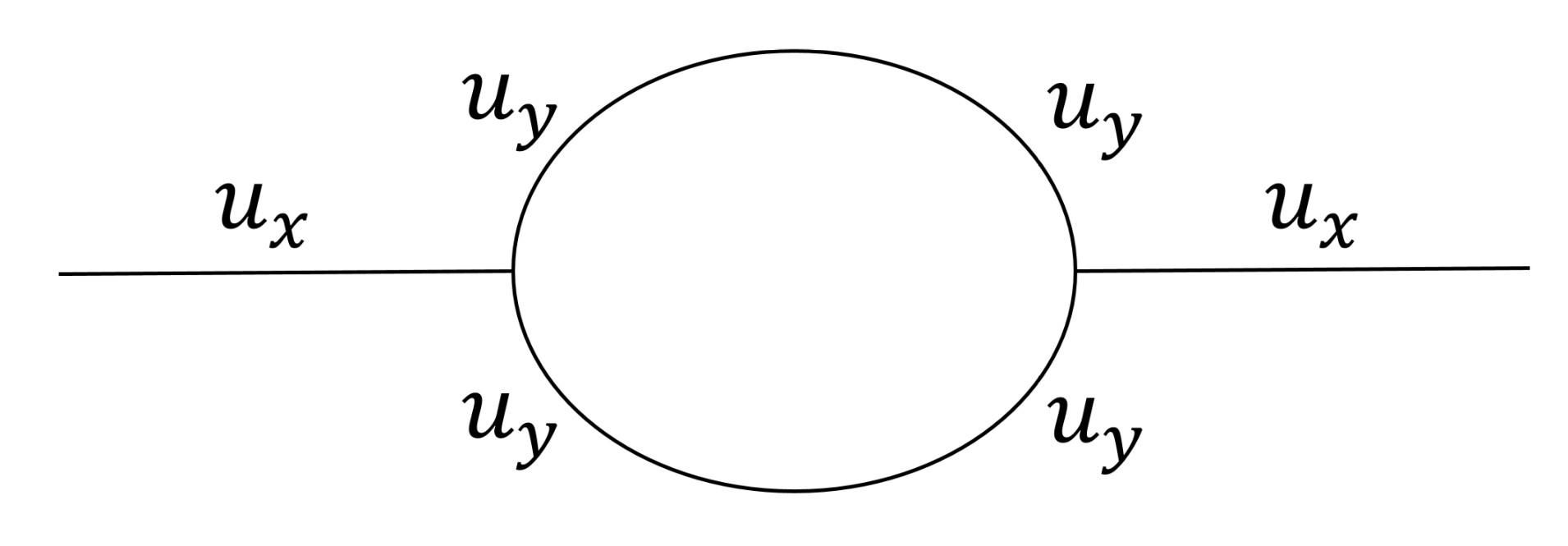}
	\end{center}
	\caption{ $|$ {\bf The one-loop graphical correction to the mass term $u_x^2$ in the Hamiltonian (\ref{Seq:H}).} %(\ref{ReHxy2}).}
		This arises from the combination of two cubic terms $u_xu_y^2$.}
	\label{CorrectionAlpha}
\end{figure}

To estimate the length scales beyond which the anharmonic terms become important,
we treat the anharmonic terms perturbatively and calculate the corrections to the harmonic terms.
These calculations can be illustrated by Feynman diagrams. For example, the correction to the mass term
is illustrated in Fig.~\ref{CorrectionAlpha}, which leads to the correction to the coefficient $\alpha$:
\beqn
\delta \alpha &=& \frac{\alpha^2}{(2\pi)^3 v_0^2} \int \dd^3 \bq {D^2\left(2\alpha q_z^2 +gq_x^2
	\right)^2\over \left[\left(\mu_x q_x^2+\mu_{\rm T}^0 q_z^2\right) \left(2\alpha q_z^2+gq_x^2\right)
	+2g\alpha q_y^2 \right]^2}
\label{eq:Sxi}
\\
&=&\frac{\alpha^{3/2}D^2}{16\sqrt{2}\pi^2 v_0^2 g^{1/2}} \int \dd q_x \int\dd q_z {\left( 2\alpha q_z^2 +gq_x^2
	\right)^{1/2}\over \left(\mu_x q_x^2+\mu_{\rm T}^0 q_z^2\right)^{3/2} }
\ ,
\eeqn
where we have integrated out $q_y$ from $-\infty$ to $\infty$.

The non-linear length $\xi_x$ in the $x$-direction is determined by the condition that, for  a system of linear extent $\xi_x$ in the $x$-direction and  infinite in the $y$ and $z$ directions, this correction (\ref{eq:Sxi}) exactly equals the bare value of $\alpha$. This leads to the condition
\beqn
\alpha =\frac{\alpha^{3/2}D^2}{4\sqrt{2}\pi^2 v_0^2 g^{1/2}} \int_{\xi_x^{-1}}^{\infty} \dd q_x \int_0^\infty\dd q_z {\left( 2\alpha q_z^2 +gq_x^2
	\right)^{1/2}\over \left(\mu_x q_x^2+\mu_{\rm T}^0 q_z^2\right)^{3/2} }\exp\left[-\left({ q_{\perp}\over\Lambda}\right)^2\right]
\ ,
\label{sxi2}
\eeqn
where
we have introduced a smooth ultraviolet cutoff $\Lambda$
through the Gaussian factor  $\exp\left[-\left({ q_{\perp}\over\Lambda}\right)^2\right]$ with $q_\perp \equiv \sqrt{q_x^2+q_z^2}$, and a factor of $4$ arises from our restriction of the integral to the first quadrant, a restriction that is convenient in the following.

To proceed further we switch to polar coordinates:
$q_x= q_{\perp}\cos \theta$, $q_z = q_{\perp}\sin \theta$; (\ref{sxi2}) then reduces to
\begin{eqnarray}
1=\frac{\alpha^{1/2}D^2}{4\sqrt{2}\pi^2 v_0^2 g^{1/2}} \int_0^{\pi\over2}\dd \theta {\left( 2\alpha \sin^2\theta +g\cos^2\theta
	\right)^{1/2}\over \left(\mu_x \cos^2\theta+\mu_{\rm T}^0 \sin^2\theta\right)^{3/2} }\int_{q_m(\theta)}^{\infty} {\dd q_{\perp} \over q_{\perp}}\exp\left[-\left({ q_{ \perp}\over\Lambda}\right)^2\right]
\  ,
\label{sxi3}
%	=\frac{\alpha^{3/2}D^2}{16\sqrt{2}\pi^2 v_0^2 g^{1/2}} A(\alpha, g, \mu_x, \mu_\perp)  \ln (L \Lambda)\,,
\end{eqnarray}
where we've defined
\beq
q_m(\theta)\equiv \xi_x^{-1}\sec\theta \ .
\label{qmdef}
\eeq

%\beq
%\frac{\alpha^{3/2}D^2}{16\sqrt{2}\pi^2 v_0^2 g^{1/2}} A(\alpha, g, \mu_x, \mu_\perp)  \ln (L \Lambda)
%\eeq
%where\beq A(\alpha, g, \mu_x, \mu_\perp)\equiv \int_0^{2\pi} \dd \theta {\left( 2\alpha \sin^2 \theta +g\cos^2\theta	\right)^{1/2}\over \left(\mu_x \cos^2\theta+\mu_{\rm T}^0 \sin^2\theta\right)^{3/2} }\ .\eeq

It is straightforward to evaluate the integral over $q_{\perp}$ in this expression for $q_m\ll\Lambda$, which will always be the case, for all angles $\theta$, when $\Lambda\xi_x\gg1$, as we will verify {\it a posteriori} that it is for small noise strength $D$. We find
\begin{eqnarray}
\int_{q_m(\theta)}^{\infty} {\dd q_{\perp} \over q_{\perp}}\exp\left[-\left({ q_{ \perp}\over\Lambda}\right)^2\right]&&=\ln\left({\Lambda\over q_m(\theta)}\right)-{C\over2}+\cO\left[\left({q_m\over\Lambda}\right)^2\ln\left({q_m\over\Lambda}\right)\right]
\nonumber\\
&&=\ln\left({\Lambda\xi_x\cos\theta}\right)-{C\over2}+\cO\left[\left({1\over\Lambda\xi_x\cos\theta}\right)^2\ln\left({1\over\Lambda\xi_x\cos\theta}\right)\right]
\ ,
\label{qint}
%	=\frac{\alpha^{3/2}D^2}{16\sqrt{2}\pi^2 v_0^2 g^{1/2}} A(\alpha, g, \mu_x, \mu_\perp)  \ln (L \Lambda)\,,
\end{eqnarray}
where $C=0.577215664...$ is Euler's constant.
%{\cblue [LC: John, I am ashamed to admit that I don't know how you evaluate this integral. But I trust that you are right.]}{\cmag [[CFL: I don't know how to do it either, but I checked that the expression is correct using Mathematica]]}{\cred [[JT: I  integrated by parts, using ${\dd q_{\cmag \perp} \over q_{\cmag \perp}}=\dd\ln q_\perp$. The integrand you then get is integrable all the way down to $q_\perp=0$, so extending that lower limit to $0$, and changing variables to $\left({q_\perp\over\Lambda}\right)^2$ gives $\int_0^\infty e^{-x}\ln(x) dx$, which is a pure number that happens to be called Euler's constant (or, actually, minus Euler's constant), and which is known to be $-.577215...$. I'm still annoyed that Mathematica can do this, though!}

Dropping the $\cO\left[\left({1\over\Lambda\xi_x\cos\theta}\right)^2\ln\left({1\over\Lambda\xi_x\cos\theta}\right)\right]$ terms, which vanish for $\Lambda\xi_x\gg1$, equation (\ref{sxi3}) becomes
\begin{eqnarray}
1=\frac{\alpha^{1/2}D^2}{4\sqrt{2}\pi^2 v_0^2 g^{1/2}}\left[A\ln\left({\Lambda'\xi_x}\right)+G\right] \  ,
\label{sxi4}
%	=\frac{\alpha^{3/2}D^2}{16\sqrt{2}\pi^2 v_0^2 g^{1/2}} A(\alpha, g, \mu_x, \mu_\perp)  \ln (L \Lambda)\,,
\end{eqnarray}
where we've defined
\begin{eqnarray}
&&A\equiv \int_0^{\pi\over2}\dd \theta {\left( 2\alpha \sin^2\theta +g\cos^2\theta
	\right)^{1/2}\over \left(\mu_x \cos^2\theta+\mu_{\rm T}^0 \sin^2\theta\right)^{3/2} }={1\over\mu_x}\sqrt{g\over\mu_{\rm T}^0} f(\g) \nonumber\\
&&G\equiv \int_0^{\pi\over2}\dd \theta {\left( 2\alpha \sin^2\theta +g\cos^2\theta
	\right)^{1/2}\over \left(\mu_x \cos^2\theta+\mu_{\rm T}^0 \sin^2\theta\right)^{3/2} }\ln\left({\cos\theta}\right)= {1\over\mu_x}\sqrt{g\over\mu_{\rm T}^0} h(\g, \Upsilon)\,,
\label{AGdef}
\end{eqnarray}
\begin{eqnarray}
f(\g)\equiv\left\{
\begin{array}{ll}
\sqrt{\g}E(\sqrt{1-\g^{-1}})\,\,\,\,\,\,\,\,\,,
&\g>1\,,
\\ \\
E(\sqrt{1-\g})\,\,\,\,\,\,\,\,\,\,\,\,\,\,\,\,\,\,,
&\g<1\,,
\end{array}\right.
\label{fdef}
\end{eqnarray}
with $E(k)$ the elliptical integral of the second kind,
\begin{eqnarray}
h(\g,\Upsilon)\equiv -{1\over2}\int_0^{\pi\over2}\dd \phi \cos\phi \sqrt{1+\g\tan^2\phi}\ln\left(1+\Upsilon\tan^2\phi\right) \label{hdef}
\end{eqnarray}
and
\beq
\g\equiv{2\alpha\mu_x\over g\mu_{\rm T}^0} \,\ , \,\,\,\, \Upsilon\equiv{\mu_x\over \mu_{\rm T}^0} \,\ , \,\,\,\,  \Lambda'\equiv\Lambda e^{-C/2} \,\,.
\label{lambda'def}
\eeq
Solving equation(\ref{sxi4}) for $\xi_x$ gives
\begin{eqnarray}
\xi_x=\Lambda^{-1}e^{C\over2}\exp\left[4\sqrt{2}\pi^2{v_0^2\mu_x\over D^2 f(\gamma)}\sqrt{\mu_{\rm T}^0\over\alpha}-{h(\g,\Upsilon)\over f(\g)}\right]\,.
\label{sxifinal}
\end{eqnarray}

We can now calculate
the non-linear length $\xi_z$ in the $z$  direction in precisely the same way; that is, by calculating the correction to $\alpha$ in  a system of linear extent $\xi_z$ in the $z$-direction and  infinite in the $x$ and $y$ directions. We now obtain
\beqn
\alpha =\frac{\alpha^{3/2}D^2}{4\sqrt{2}\pi^2 v_0^2 g^{1/2}} \int_{\xi_z^{-1}}^{\infty} \dd q_z \int_0^\infty\dd q_x {\left( 2\alpha q_z^2 +gq_x^2
	\right)^{1/2}\over \left(\mu_x q_x^2+\mu_{\rm T}^0 q_z^2\right)^{3/2} }\exp\left[-\left({ q_{\perp}\over\Lambda}\right)^2\right]
\ ,
\label{szi}
\eeqn
which can again be evaluated by switching to polar coordinates:
$q_x= q_{\perp}\cos \theta$, $q_z = q_{\perp}\sin \theta$; (\ref{sxi2}), which  gives
\begin{eqnarray}
1=\frac{\alpha^{1/2}D^2}{4\sqrt{2}\pi^2 v_0^2 g^{1/2}} \int_0^{\pi\over2}\dd \theta {\left( 2\alpha \sin^2\theta +g\cos^2\theta
	\right)^{1/2}\over \left(\mu_x \cos^2\theta+\mu_{\rm T}^0 \sin^2\theta\right)^{3/2} }\int_{q_{mz}(\theta)}^{\infty} {\dd q_{\perp} \over q_{\perp}}\exp\left[-\left({ q_{ \perp}\over\Lambda}\right)^2\right]
\  .
\label{szi2}
%	=\frac{\alpha^{3/2}D^2}{16\sqrt{2}\pi^2 v_0^2 g^{1/2}} A(\alpha, g, \mu_x, \mu_\perp)  \ln (L \Lambda)\,,
\end{eqnarray}
The only change from our calculation of $\xi_x$ is that the lower cutoff $q_{mz}(\theta)$ is now given by
\beq
q_{mz}(\theta)\equiv \xi_z^{-1}\csc\theta \ .
\label{qmzdef}
\eeq

Proceeding exactly as before, we thereby find that $\xi_z$ is determined by the condition:
\begin{eqnarray}
1=\frac{\alpha^{1/2}D^2}{4\sqrt{2}\pi^2 v_0^2 g^{1/2}}\left[A\ln\left({\Lambda'\xi_z}\right)+G_2\right] \  ,
\label{szi3}
%	=\frac{\alpha^{3/2}D^2}{16\sqrt{2}\pi^2 v_0^2 g^{1/2}} A(\alpha, g, \mu_x, \mu_\perp)  \ln (L \Lambda)\,,
\end{eqnarray}
where all symbols are as defined earlier, and
\beq
G_2\equiv \int_0^{\pi\over2}\dd \theta {\left( 2\alpha \sin^2\theta +g\cos^2\theta
	\right)^{1/2}\over \left(\mu_x \cos^2\theta+\mu_{\rm T}^0 \sin^2\theta\right)^{3/2} }\ln\left({\sin\theta}\right) \ .
\label{G2def}
\eeq
The quickest way to calculate $\xi_z$ is simply to
subtract equation (\ref{szi3}) from equation (\ref{sxi4}); this gives
\beq
A\ln\left({\xi_x\over\xi_z}\right)+G-G_2=A\ln\left({\xi_x\over\xi_z}\right)-\int_0^{\pi\over2}\dd \theta {\left( 2\alpha \sin^2\theta +g\cos^2\theta
	\right)^{1/2}\over \left(\mu_x \cos^2\theta+\mu_{\rm T}^0 \sin^2\theta\right)^{3/2} }\ln\left({\tan\theta}\right) =0 \ ,
\label{sxiszi}
\eeq
which can obviously be solved for the natural logarithm of the ratio $\xi_x/\xi_z$:
\beq
\ln\left({\xi_x\over\xi_z}\right)={1\over A}\int_0^{\pi\over2}\dd \theta {\left( 2\alpha \sin^2\theta +g\cos^2\theta
	\right)^{1/2}\over \left(\mu_x \cos^2\theta+\mu_{\rm T}^0 \sin^2\theta\right)^{3/2} }\ln\left({\tan\theta}\right) \ .
\label{sxiszi2}
\eeq

By changing variable of integration from $\theta$ to $u\equiv\tan\theta$, and then changing variables again to $\phi$ defined by $u=\sqrt{\mu_x\over\mu_{\rm T}^0} \tan\phi$ gives
\beq
\int_0^{\pi\over2}\dd \theta {\left( 2\alpha \sin^2\theta +g\cos^2\theta
	\right)^{1/2}\over \left(\mu_x \cos^2\theta+\mu_{\rm T}^0 \sin^2\theta\right)^{3/2} }\ln\left({\tan\theta}\right)={A\over2}\ln\left({\mu_x\over\mu_{\rm T}^0}\right)+{1\over\mu_x}\sqrt{g\over\mu_{\rm T}^0}\int_0^{\pi\over2}\dd \phi \sqrt{\g \sin^2\phi+\cos^2\phi}\ln\left({\tan\phi}\right) \ .
\label{sxiszi3}
\eeq
Using this in (\ref{sxiszi2}), and using our earlier result (\ref{AGdef}) for A, we obtain
\beq
\ln\left({\xi_x\over\xi_z}\right)={1\over2}\ln\left({\mu_x\over\mu_{\rm T}^0}\right)+{f_2(\g)\over f(\g)} \ ,
\label{sxiszi4}
\eeq
where we've defined
\beq
f_2(\g)\equiv\int_0^{\pi\over2}\dd \phi \sqrt{\g \sin^2\phi+\cos^2\phi}\ln\left({\tan{\phi}}\right) \ .
\label{f2def}
\eeq
It is clear by inspection that $f_2$ is $\cO(1)$ when $\g=\cO(1)$.
It is also straightforward to show that when $\g\gg1$, $f_2(\g)\approx\sqrt{\g}\ln2$.  Inspection of (\ref{fdef}) shows that $f(\g)$ is always greater than $1$ , and that $f(\g)=\cO(1)$ when $\g=\cO(1)$. In addition, when $\g\gg1$, $f(\g)\approx\sqrt{\g}$. Putting this all together, we see that, {\it whatever} the value of $\g$, the ratio ${f_2(\g)\over f(\g)}=\cO(1)$.
%{\cblue [LC: I tested this ratio numerically and suspect  $|{f_2(\g)\over f(\g)}|<\ln 2$, which makes (\ref{sxiszi6}) more convincing.]}
Hence, equation (\ref{sxiszi4}) implies
\beq
\ln\left({\xi_x\over\xi_z}\right)={1\over2}\ln\left({\mu_x\over\mu_{\rm T}^0}\right)+\cO(1) \ ,
\label{sxiszi5}
\eeq
which in turn implies
\beq
{\xi_x\over\xi_z}=\sqrt{\mu_x\over\mu_{\rm T}^0}\times\cO(1) \ .
\label{sxiszi6}
\eeq
This result is, of course, exactly what we would have gotten if we had assumed that the two pieces of the factor $\mu_x q_x^2+\mu_{\rm T}^0 q_z^2$ in the propagator are comparable to each other when $q_x=\xi_x^{-1}$ and $q_z=\xi_z^{-1}$.

We can now apply the above analysis to a system of finite extent $\xi_y$ in the $y$-direction and infinite in the $x$ and $z$-directions.
The condition determining $\xi_y$ is:
\beqn
1&=& \frac{\alpha}{(2\pi)^3 v_0^2} \int \dd^3 \bq {D^2\left(2\alpha q_z^2 +gq_x^2
	\right)^2\over \left[\left(\mu_x q_x^2+\mu_{\rm T}^0 q_z^2\right) \left(2\alpha q_z^2+gq_x^2\right)
	+2g\alpha q_y^2 \right]^2}
\nonumber \\
%\label{eq:Sxi}
&=&\frac{\alpha D^2}{\pi^3 v_0^2} \int_{\xi_y^{-1}}^\infty \dd q_y \int_0^{\pi\over2}\dd \theta \int_0^\infty \dd q_\perp{\left( 2\alpha \sin^2\theta +g\cos^2\theta
	\right)^2 q_\perp^5\over \left[\left( 2\alpha \sin^2\theta +g\cos^2\theta
	\right) \left(\mu_x \cos^2\theta+\mu_{\rm T}^0 \sin^2\theta\right)q_\perp^4+2\alpha g q_y^2 \right]^2}\exp\left[-\left({ q_{\perp}\over\Lambda}\right)^2\right]\nonumber\\
\label{syi}
\eeqn
where we've now defined our polar coordinates $q_\perp$ and $\theta$ via
$q_x= q_\perp\cos \theta$, $q_z = q_\perp\sin \theta$.

The integral over $q_\perp$ in this expression converges as $q_\perp\to\infty$ even without the Gaussian ultraviolet cutoff. This implies that the integral itself will be  insensitive to the ultraviolet cutoff $\Lambda$ provided that the $q_\perp^4$ term in the denominator of (\ref{syi}) dominates the $q_y$ term even for $q_\perp\sim\Lambda$. Thus we can throw out the Gaussian factor in that integral whenever $2\alpha g q_y^2\ll \left( 2\alpha \sin^2\theta +g\cos^2\theta
\right) \left(\mu_x \cos^2\theta+\mu_{\rm T}^0 \sin^2\theta\right)\Lambda^4$. That is, we can neglect that cutoff for
\beq
q_y\ll\Lambda_y(\theta)\equiv\sqrt{\left( 2\alpha \sin^2\theta +g\cos^2\theta
	\right) \left(\mu_x \cos^2\theta+\mu_{\rm T}^0 \sin^2\theta\right)\over2\alpha g}\Lambda^2 \ .
\label{UVydef}
\eeq

We'll show in a moment that, in this regime, the integral over $q_\perp$ scales like $q_y^{-1}$.
On the other hand, for $q_y\gg\Lambda_y(\theta)$, the $q_\perp$ integral converges for $q_\perp\sim\Lambda$, at which values of $q_\perp$ the denominator of (\ref{syi}) is dominated by the $q_y^2$ term. In this case,  the integral over $q_\perp$ scales like $q_y^{-4}$.
The integral of the latter over $q_y$ then converges rapidly as $q_y\to\infty$. This implies that
$\Lambda_y(\theta)$ acts as an effective ultraviolet cutoff on the integral over $q_y$. We can therefore approximate  (\ref{syi}) by
\beqn
1=\frac{\alpha D^2}{\pi^3 v_0^2} \int_0^{\pi\over2}\dd \theta \int_{\xi_y^{-1}}^{\Lambda_y(\theta)} \dd q_y  \int_0^\infty \dd q_\perp{\left( 2\alpha \sin^2\theta +g\cos^2\theta
	\right)^2 q_\perp^5\over \left[\left( 2\alpha \sin^2\theta +g\cos^2\theta
	\right) \left(\mu_x \cos^2\theta+\mu_{\rm T}^0 \sin^2\theta\right)q_\perp^4+2\alpha g q_y^2 \right]^2} \ .
\label{syi2}
\eeqn
Note that although our argument for the effective ultraviolet cutoff $\Lambda_y$ (\ref{UVydef}) is rather rough, the equation (\ref{syi2}) should be quite exact, due to the weakness of the dependence of the integral over $q_y$ on that ultraviolet cutoff (that is, the fact that it depends only logarithmically on that cutoff).

The elementary integral over $q_\perp$ is
\beq
\int_0^\infty \dd q_\perp{\left( 2\alpha \sin^2\theta +g\cos^2\theta
	\right)^2 q_\perp^5\over \left[\left( 2\alpha \sin^2\theta +g\cos^2\theta
	\right) \left(\mu_x \cos^2\theta+\mu_{\rm T}^0 \sin^2\theta\right)q_\perp^4+2\alpha g q_y^2 \right]^2}={\pi\over8} {\left( g^{-1}\sin^2\theta +(2\alpha)^{-1}\cos^2\theta
	\right)^{1/2}\over \left(\mu_x \cos^2\theta+\mu_{\rm T}^0 \sin^2\theta\right)^{3/2} q_y } \ .
\label{qpint}
\eeq
Inserting this into (\ref{syi2}) and performing the trivial integral over $q_y$ leads to
\begin{eqnarray}
1=\frac{\alpha^{1/2}D^2}{8\sqrt{2}\pi^2 v_0^2 g^{1/2}} \int_0^{\pi\over2}\dd \theta {\left( 2\alpha \sin^2\theta +g\cos^2\theta
	\right)^{1/2}\over \left(\mu_x \cos^2\theta+\mu_{\rm T}^0 \sin^2\theta\right)^{3/2} }\ln\left({\Lambda_y(\theta)\xi_y}\right) \  ,
\label{syi3}
%	=\frac{\alpha^{3/2}D^2}{16\sqrt{2}\pi^2 v_0^2 g^{1/2}} A(\alpha, g, \mu_x, \mu_\perp)  \ln (L \Lambda)\,,
\end{eqnarray}
Using our expression (\ref{UVydef}) for $\Lambda_y(\theta)$ in this expression, we can rewrite it as
\begin{eqnarray}
1=\frac{\alpha^{1/2}D^2}{8\sqrt{2}\pi^2 v_0^2 g^{1/2}}\left[A\ln\left({\Lambda^2\xi_y}\right)+G_y\right] \  ,
\label{syi4}
%	=\frac{\alpha^{3/2}D^2}{16\sqrt{2}\pi^2 v_0^2 g^{1/2}} A(\alpha, g, \mu_x, \mu_\perp)  \ln (L \Lambda)\,,
\end{eqnarray}
where $A$ was defined in (\ref{AGdef}) and we've defined
\beq
G_y\equiv {1\over2}\int_0^{\pi\over2}\dd \theta {\left( 2\alpha \sin^2\theta +g\cos^2\theta
	\right)^{1/2}\over \left(\mu_x \cos^2\theta+\mu_{\rm T}^0 \sin^2\theta\right)^{3/2} }\ln\left[\left( g^{-1}\sin^2\theta +(2\alpha)^{-1}\cos^2\theta
\right) \left(\mu_x \cos^2\theta+\mu_{\rm T}^0 \sin^2\theta\right)\right]\,.
\label{Gydef}
\eeq
We can easily use (\ref{syi4}) to obtain a simple relation between $\xi_x$ and $\xi_y$ by  taking the ratio of equation (\ref{syi4}) to equation  (\ref{sxi4}), which gives
\beq
{\left[A\ln\left({\Lambda^2\xi_y}\right)+G_y\right]
	\over
	2\left[A\ln\left({\Lambda'\xi_x}\right)+G\right]}=1 \ .
\label{ratio}
\eeq
This in turn implies
\beq
A\ln\left({\Lambda^2\xi_y}\right)+G_y
=
2A\ln\left({\Lambda'\xi_x}\right)+2G \ .
\label{sxisyi}
\eeq
Gathering the logarithmic terms on one side of this expression, and the constant terms on the other, gives
\beq
A\ln\left({\Lambda^2\xi_y\over\left(\Lambda'\xi_x\right)^2}\right)
=
2G-G_y \ ,
\label{sxisyi2}
\eeq
which can be solved for $\xi_y$:
\beq
\xi_y=\xi_x^2\exp\left({2G-G_y\over A}-C\right) \ ,
\label{syisol}
\eeq
where we have used (\ref{lambda'def}) for $\Lambda'$.
We can further simplify this expression by writing the combination $2G-G_y$ as a single integral over $\theta$:
\begin{eqnarray}
2G-G_y&=&\int_0^{\pi\over2}\dd \theta {\left( 2\alpha \sin^2\theta +g\cos^2\theta
	\right)^{1/2}\over \left(\mu_x \cos^2\theta+\mu_{\rm T}^0 \sin^2\theta\right)^{3/2} }\left(2\ln(\cos\theta)-{1\over2}\ln\left[\left( {1\over g}\sin^2\theta +{1\over 2\alpha}\cos^2\theta\right) \left(\mu_x \cos^2\theta+\mu_{\rm T}^0 \sin^2\theta\right)\right]\right) \nonumber \\
&=& -{1\over2}\int_0^{\pi\over2}\dd \theta {\left( 2\alpha \sin^2\theta +g\cos^2\theta\right)^{1/2}\over \left(\mu_x \cos^2\theta+\mu_{\rm T}^0 \sin^2\theta\right)^{3/2} }\left(\ln\left[\left( {1\over g}\tan^2\theta +{1\over 2\alpha}
\right) \left(\mu_x +\mu_{\rm T}^0 \tan^2\theta\right)\right]\right)
\,.
\label{ggy}
\end{eqnarray}
Pulling a factor of ${\mu_x\over2\alpha}$ out of the argument of the  logarithm
in this expression, and using our definition (\ref{AGdef}) of $A$, we can rewrite this as
\begin{eqnarray}
2G-G_y= -{A\over2} \ln\left({\mu_x\over2\alpha}\right)-{1\over2}\int_0^{\pi\over2}\dd \theta {\left( 2\alpha \sin^2\theta +g\cos^2\theta\right)^{1/2}\over \left(\mu_x \cos^2\theta+\mu_{\rm T}^0 \sin^2\theta\right)^{3/2} }\left(\ln\left[\left( {2\alpha\over g}\tan^2\theta +1
\right) \left(1 +{\mu_{\rm T}^0\over\mu_x} \tan^2\theta\right)\right]\right)
\,.
\label{ggy2}
\end{eqnarray}
%Now introducing the new variable of integration variable  $u\equiv\tan\theta$, followed by a change to another variable of integration $w\equiv\sqrt{\mu_x\over\mu_{\rm T}^0} u$,
Now again by changing variable of integration from $\theta$ to $u\equiv\tan\theta$, and then changing variable to $\phi$ defined by $u=\sqrt{\mu_x\over\mu_{\rm T}^0} \tan\phi$, we obtain, using our definition of $A$ (\ref{AGdef}) again,
\begin{eqnarray}
2G-G_y= -{A\over2}\left( \ln\left({\mu_x\over2\alpha}\right)+\Psi(\g)\right)
\, ,
\label{ggy3}
\end{eqnarray}
where
we've defined
\begin{eqnarray}
\Psi(\g)= {1\over f(\g)}\int_0^{\pi\over2}\dd \phi \ \ {\cos\phi\sqrt{1+\g \tan^2\phi}}\ln\left(1+\g\tan^2\phi\over\cos^2\phi\right)
\, ,
\label{Psidef}
\end{eqnarray}
where $f(\g)$ was defined in (\ref{fdef}), and $\g$ was defined in (\ref{lambda'def}). Using the limits (\ref{fdef}) on $f(\g)$ and this expression (\ref{Psidef}), it is straightforward to show that the limiting behaviors of $\Psi$ are
\begin{eqnarray}
\Psi(\g)=\left\{
\begin{array}{ll}
\cO(1)\,\,\,\,\,\,\,\,\,,
&\g=\cO(1)\,,
\\ \\
\ln(\g)+\cO(1)\,\,\,\,\,\,\,\,\,\,\,\,\,\,\,\,\,\,,
&\g\gg1\,,
\end{array}\right.
\label{Psilim}
\end{eqnarray}
With the result (\ref{ggy3}) in hand,  we can rewrite (\ref{syisol}) as
\beq
\xi_y={\xi_x^2\over\lambda_x}\exp\left(-\frac{1}{2}\Psi(\g)+C\right) \ ,
\label{syin}
\eeq
where we've defined
\beq
\lambda_x\equiv\sqrt{\mu_x\over2\alpha} \ .
\eeq
Using the limiting behaviors  (\ref{Psilim})
of $\Psi$ that we just derived, we can obtain the limiting behaviors of $\xi_y$:
\begin{eqnarray}
\xi_y=\left\{
\begin{array}{ll}
\xi_y={\xi_x^2\over\lambda_x}\times\cO(1)\,\,\,\,\,\,\,\,\,,
&\g=\cO(1)\,,
\\ \\
{\xi_x^2\over\lambda_x\sqrt{\g}}\times\cO(1)\,\,\,\,\,\,\,\,\,\,\,\,\,\,\,\,\,\,,
&\g\gg1\, .
\end{array}\right.
\label{syilim}
\end{eqnarray}
We can conveniently summarize these two limiting behaviors with a single interpolation formula:
\beq
\xi_y={\xi_x^2\over\lambda_x\sqrt{1+\g}}\times\cO(1) \ .
\label{syin}
\eeq
Again this result is exactly what we would have gotten if we had assumed that the two pieces $\left(\mu_x q_x^2+\mu_{\rm T}^0 q_z^2\right)\left(2\alpha q_z^2+gq_x^2\right)$ and $2g\alpha q_z^2$ in the denominator in the propagator are comparable to each other when $q_x=\xi_x^{-1}$, $q_z=\xi_z^{-1}$, and $q_y=\xi_y^{-1}$.

Note that all three non-linear lengths diverge exponentially (i.e., like $\exp\left[{{\rm constant}\over D^2}\right]$) as the noise strength $D\rightarrow0$. This strong divergence implies that, in systems with weak noise (small $D$), these three non-linear lengths could become astronomically large. In such systems, the non-linear effects we've described in this paper would be undetectable in any realistically sized flock. In this case, our result (\ref{uuSI1}) would hold with the parameters $\alpha$, and $\mu_{x,z}$ simply being constants, rather than the logarithmically diverging or vanishing functions of $q$ that they become for $q$ smaller than $\xi_{x,y,z}$.

In such low noise systems, the logarithms in the real space correlation functions equation (1) of the main text also disappear, leaving only the power law dependences on $x$,$y$, and $z$ given there.

%\beqn
%\alpha&=&\frac{\alpha^{3/2}D^2}{16\sqrt{2}\pi^2 v_0^2 g^{1/2}} A(\alpha, g, \mu_x, \mu_\perp)  \ln (\xi_{xz}\Lambda)
%\\\xi_{xz} &=& \Lambda^{-1} \exp \left[\frac{16\sqrt{2}\pi^2 v_0^2 g^{1/2}}{\alpha^{1/2}D^2  A(\alpha, g, \mu_x, \mu_\perp)}\right] \ .\eeqn

%
%
%
%
%\begin{eqnarray}
%\delta\alpha&=&{\alpha^2\over (2\pi)^3v_0^2}\int d^3p\ \ {D(2\alpha p_z^2+gp_x^2)^2
%	\over \left[\mu (2\alpha p_z^2+gp_x^2)p^2+2\alpha gp_y^2\right]^2}\nonumber\\
%&=&{D\alpha^{3\over 2}\over 16\sqrt{2}\pi^2v_0^2g^{1\over 2}\mu^{3\over 2}}
%\int dp_xdp_z\ \ {\sqrt{2\alpha p_z^2+gp_x^2}\over q_{\perp}^3}\nonumber\\
%&=&{D\alpha^{3\over 2}\Gamma\left(2\alpha/g\right)\over 16\sqrt{2}\pi^2v_0^2\mu^{3\over 2}}\ln\left(L\Lambda\right)\,,\label{ReDelta_alpha}
%\end{eqnarray}
%where
%\begin{eqnarray}
%\Gamma(\tau)=\int_0^{2\pi} d\theta \sqrt{\cos^2\theta+\tau\sin^2\theta}\,,
%\end{eqnarray}
%$\Lambda$ is the ultraviolet cutoff, and $L$ is the linear extension of the system. In the thermodynamic limit $L\to\infty$, $\delta\alpha$ diverges logarithmically. This implies the harmonic theory breaks down and the anharmonic terms in (\ref{ReHxy2}) are marginally relevant, which is consistent with our earlier conclusion that the non-linear terms whose coefficients are proportional to $\alpha$ in (\ref{ReFullEOM1}) are ``marginal".

%%%%%%%%%%%%%%%%%%%
%%%%%%%%%%%%%%%%%%%%%%%
%
%

%\sout{Alternatively, when doing the calculation in (\ref{eq:Sxi}), we can do a full range integral over $q_{x,z}$ first %(\ref{ReDelta_alpha})
%and then a finite range integral on $q_y$ with the upper bound $\Lambda$ and the lower bound $1/L$. This gives
%another expression for $\delta\alpha$. Then equating $\delta\alpha$ to $\alpha$ gives the crossover length along
%$y$:}
%{\cred JT cut an equation here.}
%	\begin{eqnarray}\xi_y=\Lambda\left(\xi_{xz}\right)^2=\Lambda^{-1}\exp\left[32\sqrt{2}\pi^2v_0^2\mu^{3\over 2}\over D\alpha^{1\over 2}\Gamma\left(2\alpha\over g\right)\right]\,.\end{eqnarray}
%{\cmag
%\beq
%\xi_{y} = \Lambda^{-1} \exp \left[\frac{64\sqrt{2}\pi^2 v_0^2 g^{1/2}}{\alpha^{1/2}D^2  A(\alpha, g, \mu_x, \mu_\perp)}\right] =\Lambda^{-3} \xi_{xz}^4
%\ .
%\eeq
%}

%\noindent{\cmag [[CFL: I found $\xi_{y} \sim \xi_z^{4}$ instead of $\xi_{y} \sim \xi_z^{2}$.]]}
%
%\noindent{\cblue [LC: I did the calculation again and confirmed that the above result is correct.]}
%	%\begin{eqnarray}
%%\xi_y^s=\Lambda\left(\xi_{xz}^s\right)^2=\Lambda^{-1}\exp\left[32\sqrt{2}\pi^2v_0^2\mu^{3\over 2}\over D\alpha^{1\over 2}\Gamma\left(2\alpha\over g\right)\right]\,.
%%\end{eqnarray}
%
%{\cred JT: Well, as you guys can see, I got a different answer from both of you; have a look at what I've done above and let me know what you think.}

\section{Calculation of the real-space equal-time correlation functions}
Since we have obtained the equal-time correlation function in the momentum space, the equal-time correlation
function in real space can be calculated through inverse Fourier transformation:
%\begin{widetext}
	\begin{eqnarray}
	C(\br)\equiv\langle\bu(\br)\cdot\bu({\bf 0})\rangle
	&=&\int {d^3q\over(2\pi)^3}\int {d^3q'\over(2\pi)^3}\langle\bu(\bq,t)\cdot\bu(\bq',t)\rangle \exp\left(i \bq\cdot\br\right)
	\nonumber\\
	&=&\int {d^3q\over(2\pi)^3}{D\left(gq_x^2+2\alpha q_z^2\right) \exp\left(i \bq\cdot\br\right)
		\over \left(\mu_xq_x^2+\mu_{\rm T}^0q_z^2\right)\left(gq_x^2+2\alpha q_z^2\right)+2g\alpha q_y^2}~.
	%\nonumber \\ &&=\int {d^3q\over(2\pi)^3}{Dq_x^2 e^{i \bq\cdot\br}
	%\over \left[\mu_xq_x^2+\mu_{\rm T}^0q_z^2\right] q_x^2+2\alpha q_y^2}~,
	\label{}
	\end{eqnarray}
%\end{widetext}
We will first do this calculation in the linear theory, treating all the coefficients as constants.
Once we get the result for the linear theory, we then take these coefficients to be length-dependent
so as to take into account anharmonic effects. The length-dependences of these coefficients are inferred
from their $q$-dependences, as given by equation (24) of the main text,  by replacing $q$ with $1\over r$.

We do the integral over $q_y$ first using complex contour techniques. This gives
\begin{eqnarray}
C(\br)%\nonumber \\ &&
={D\over8\pi^2\sqrt{2g\alpha}}\int_{-\infty}^\infty dq_x\int_{-\infty}^\infty dq_z {\sqrt{2\alpha q_z^2+gq_x^2}
	\exp\left(-\sqrt{2\alpha q_z^2+gq_x^2}\sqrt{\mu_xq_x^2+\mu_{\rm T}^0q_z^2\over2g\alpha}|y|
	+i\bq\cdot\br_\perp\right)
	\over \sqrt{\mu_xq_x^2+\mu_{\rm T}^0q_z^2}}~,
\label{SICr1}
\end{eqnarray}
where $\perp$ denotes $xz$ plane.

Now let's consider $C(x,y=0,z=0)$:
%\begin{eqnarray}
%C(x,y=z=0)%\nonumber \\ &&
%&=&{D\over8\pi^2\sqrt{2g\alpha}}\int_{-\infty}^{\infty} dq_x
%\int_{-\infty}^{\infty}dq_z {\sqrt{2\alpha q_z^2+gq_x^2} \exp\left(iq_xx-{q_{\perp}^2\over \Lambda^2}\right)
%\over \sqrt{\mu_xq_x^2+\mu_{\rm T}^0q_z^2}}\nonumber\\
%&=&{D\over8\pi^2\sqrt{g\mu_{\rm T}^0}}\int_{-\infty}^{\infty} dq_x\int_{-\infty}^{\infty}dq_z
%{\sqrt{q_z^2+q_2^2} \exp\left(iq_xx-{q_{\perp}^2\over \Lambda^2}\right)
%\over \sqrt{q_z^2+q_1^2}}\,,
%\label{Cx1}
%\end{eqnarray}
%where $q_1\equiv \sqrt{\mu_x/\mu_{\rm T}^0}|q_x|$, $q_2\equiv \sqrt{g/2\alpha}|q_x|$, and we have imposed a
%soft Gaussian cutoff $\Lambda$ on $q_{\perp}$. Note that the $q$-dependences
%of the coefficients $\mu_x$, $ \mu_{\rm T}^0$, $g$, and $\alpha$ imply that  $q_2\gg q_1$. %\sout{$q_2\gg q_1$, which is implied by the $q$-dependence
%%of the coefficients}.
%	
%	
%So far, we have treated our coefficients as constant when doing the inverse Fourier transform. We will now demonstrate using a different method why such an assumption is reasonable.
%		Go back to eqn. (\ref{Cx1}), and rewrite it as
\begin{eqnarray}
C(x,y=z=0)%\nonumber \\ &&
&=&{D\over8\pi^2}\int_{-\infty}^{\infty} dq_x \exp\left(iq_xx-{q_x^2\over \Lambda^2}\right)
\int_{-\infty}^{\infty}{dq_z\over\sqrt{2g\alpha}} {\sqrt{2\alpha q_z^2+gq_x^2} \exp\left(-{q_z^2\over \Lambda^2}\right)
	\over \sqrt{\mu_xq_x^2+\mu_{\rm T}^0q_z^2}}\,.
\label{Cx2_2}
\end{eqnarray}
%where all of the $q$-dependence of $g$, $\alpha$, and $\mu_{x,T}$ are kept in this expression.
Now, define
\begin{eqnarray}
I(q_x)&\equiv&%\nonumber \\ &&
\int_{-\infty}^{\infty}{dq_z\over\sqrt{2g\alpha}} {\sqrt{2\alpha q_z^2+gq_x^2} \exp\left(-{q_z^2\over \Lambda^2}\right)
	\over \sqrt{\mu_xq_x^2+\mu_{\rm T}^0q_z^2}}\,,
\label{Iqxdef}
\end{eqnarray}
%Even with the dependencies of $\bq$ in the coefficients,
one can see that this integral converges as $q_x\to0$. Hence, we now write
\beq
I(q_x)=I(q_x=0)+\delta I(q_x) \ ,
\label{delIdef}
\eeq
where $\delta I(q_x\to0)\to0$ by definition. %{\cred [LC: Here I removed discussions about the $q_x$-dependence of the coefficients, which I suppose John only wrote to illuminate me. We have said at the begging that the calculation will be done in a harmonic approximation first, and the anharmonic effect will be taken into account at the end. Keeping those discussion may cause confusion.]}
Putting this into (\ref{Cx2_2}), we see immediately that
%, {\it even taking the $\bq$ dependence of $g$, $\alpha$, and $\mu_{x,T}$ into account},
the $I(q_x=0)$ piece of this gives rise {\it only} to a short-ranged contribution $\propto e^{-(\Lambda x)^2}$ to $C(x)$. Hence, all of the long-ranged pieces of $C(x)$ come from $\delta I(q_x)$. In the following we will first calculate ${\pp I(q_x)/\pp q_x^2}$, and then from that result we derive $\delta I(q_x)$.
%There will be two types of terms in this differentiation: those coming from differentiating the explicit $q_x$ dependences of the integrand in (\ref{Cx2}), and those that come from differentiating the implicit $q_x$ dependences of $g$, {\cmag [[CFL: As Leiming pointed out, when using the log corrections in the MT, g does not depend on $q$, right?]]} $\alpha$, and $\mu_{x,T}$. The latter, however, are always much smaller than the former. To see that, consider taking ${d\over dx} (\ln^\alpha(x) x^\beta)$. This is, of course, ${d\over dx} (\ln^\alpha(x) x^\beta)=\alpha\ln^{\alpha-1}(x)x^{\beta-1}+\beta\ln^\alpha(x) x^{\beta-1}$, where the first term comes from differentiating the $\ln$, while the second comes from differentiating the $x^\beta$. The ratio of these two terms is ${\alpha\over\beta\ln(x)}$, which clearly vanishes as $x\to0$. we can ignore any contributions to ${\pp I(q_x)/\pp q_x^2}$ coming from
%		letting the derivative act on the implicit $q_x$ dependences of $g$, $\alpha$, and $\mu_{x,T}$; that is, for the purposes of that derivative, we can replace them with constants.

Changing the variables of integration in (\ref{Iqxdef}) from $q_z$ to $Q_z$ defined via $q_z\equiv\sqrt{\mu_x\over\mu_{\rm T}^0}Q_z$, we get
\beq
{\pp \delta I(q_x)\over\pp q_x^2}=%{1\over\mu_{\rm T}^0}\sqrt{\mu_x\over g}\Gamma I_1(q_x)
\frac{1}{2 \mu_{\rm T}^0} \sqrt{\frac{\mu_x}{g}} \left[ \Gamma I_1(q_x)-I_2(q_x)\right]
\label{ppI1}
\eeq
where
\beq
I_1(q_x)\equiv \int_{-\infty}^{\infty}dQ_z { \exp\left(-{Q_z^2\over \Lambda_z^2}\right)
	\over \sqrt{(q_x^2+Q_z^2)(\Gamma q_x^2+Q_z^2)}}\ ,
\label{I1def}
\eeq
and
\beq
I_2(q_x)\equiv \int_{-\infty}^{\infty}dQ_z  \exp\left(-{Q_z^2\over \Lambda_z^2}\right)
\sqrt{(\Gamma q_x^2+Q_z^2)\over(q_x^2+Q_z^2)^3}\ ,
\label{I2def}
\eeq
where we've defined $\Gamma\equiv{g\mu_{\rm T}^0\over 2\alpha\mu_x}$ and $\Lambda_z\equiv\Lambda\sqrt{\mu_{\rm T}^0\over\mu_x}$.

We will first calculate $I_1$. Since the integral converges rapidly before $Q_z$ is comparable to $\Lambda_z$, we can ignore the Gaussian cutoff.		
By a change of variable $Q_z =|q_x| u$, we have
\beq
I_1 \approx \frac{2}{|q_x|} \int_0^\infty \frac{d u}{\sqrt{(\Gamma +u^2)(1+u^2)}}
\ .
\eeq
The anomalous $q$-dependence of $\alpha$, $\mu_x$, and $\mu_{\rm T}$ given by equation (24) of the main text implies that  $\Gamma = 2g\mu_{\rm T}^0/(\alpha\mu_x )\sim   |\ln q|^{1/2}\to\infty$ as $q\rightarrow 0$.  We can exploit this fact to
split the integral into two  easily approximated parts by introducing a constant $B$ such that $1\ll B\ll \sqrt{\Gamma}$:
\beqn
I_1 &\approx& \frac{2}{|q_x|} \left[\int_0^{B} \frac{d u}{\sqrt{\Gamma (1+u^2)}}+\int_{B}^\infty \frac{d u}{u\sqrt{(\Gamma +u^2)}}\right]
\\
&=& \frac{2}{\sqrt{\Gamma}|q_x|} \left[{\rm arcsinh} (B) +{\rm arcsinh} \left(\frac{\sqrt{\Gamma}}{B}\right)\right]
\\
\label{eq:SI_I1}
&\approx&\frac{2}{\sqrt{\Gamma}|q_x|} \left[\ln (2B) +\ln \left(\frac{2\sqrt{\Gamma}}{B}\right)\right]
\\
&=&\frac{\ln (16 \Gamma)}{\sqrt{\Gamma}|q_x|}
\ ,
\eeqn
where we have used the fact that $B\gg 1$ and $\sqrt{\Gamma } \gg B$ and used the leading asymptotic expression for the ${\rm arcsinh}$ functions in (\ref{eq:SI_I1}).

We now focus on $I_2$, again with the Gaussian cutoff ignored. By the same change of variable  $Q_z =|q_x| u$, we get
\beqn
I_2(q_x)&\approx&\frac{2}{|q_x|} \int_{0}^{\infty}du
\sqrt{\Gamma +u^2\over(1+u^2)^3}
\\
&\approx& \frac{2}{|q_x|} \left[\int_0^{B}du \sqrt{\Gamma \over(1+u^2)^3}+\int_{B}^\infty du \frac{\sqrt{\Gamma +u^2}}{ u^3}\right]
\\
&=&  \frac{2}{|q_x|} \left[\frac{\sqrt{\Gamma}(B+B^3) }{(1+B^2)^{3/2}}+
\frac{1}{2}\left(\frac{\sqrt{B^2+\Gamma}}{B^2} + {{\rm arcsinh} (B/\sqrt{\Gamma})\over\sqrt{\Gamma}} \right)
\right]
\\
&\approx & \frac{2 \sqrt{\Gamma}}{|q_x|}
\eeqn
where we have used the fact that $B \gg 1$ in the last approximation.

Substituting the expressions for $I_1$ and $I_2$ back into (\ref{ppI1}), we see that, once the wavevector dependences of $\mu_{\rm T}^0$, $\mu_x$, and $\alpha$ are taken into account,  $\Gamma I_1\approx{1\over|q_x|}\sqrt{\Gamma}\ln\Gamma\gg I_2\approx{1\over|q_x|}\sqrt{\Gamma}$ as $\bq\to0$, since, as noted earlier, $\Gamma\to\infty$ in that limit. Therefore,  the $I_1$ term in (\ref{ppI1}) dominates, so 
\beq
{\pp \delta I(q_x)\over\pp q_x^2}\approx
\frac{1}{2 \mu_{\rm T}^0} \sqrt{\frac{\mu_x \Gamma }{g}} \frac{\ln (16 \Gamma)}{|q_x|} = \frac{1}{2 \sqrt{2 \alpha \mu_{\rm T}^0}} \frac{\ln (16 \Gamma)}{|q_x|}
\ .
\eeq
This is easily integrated (keeping in mind that the variable of integration is $q_x^2$, {\it not} $q_x$) to  obtain
%$I(q_x)$ by integrating $\delta I(q_x)$ with respect to $q_x^2$ to obtain
\beq
\delta I(q_x) \approx \frac{\ln (16 \Gamma)}{ \sqrt{2 \alpha \mu_{\rm T}^0}} |q_x|
\ .
\eeq
%Note that in the above, we have ignored the logarithmic dependence on $\bq$ in the coefficients when performing the integration for the same reason that we ignore that dependency when performing a differentiation.

Substituting this expression back into (\ref{Cx2_2}) we get
%we see that we have recovered \eq (\ref{3 regimes}), except for the constant term $\sqrt{\pi}\Lambda$, which shows does not affect the large distance behaviour.
\begin{eqnarray}
C(x,y=z=0)%\nonumber\\
&=&{D\ln\left(g\mu_{\rm T}^0\over 8\alpha\mu_x\right)\over 8\pi^2\sqrt{2\alpha\mu_{\rm T}^0}}
\int_{-\infty}^{\infty} dq_x q_x\cos\left(q_xx\right)\exp\left(-{q_x^2\over\Lambda^2}\right)\nonumber\\
&=&{D\ln\left(g\mu_{\rm T}^0\over 2\alpha\mu_x\right)\over 4\pi^2\sqrt{2\alpha\mu_{\rm T}^0}}{1\over x^2}
\int_0^{\infty} dQ_x Q_x\cos Q_x \exp\left(-{Q_x^2\over\Lambda^2x^2}\right)\,,
\label{SICx1}
\end{eqnarray}
where in the second equality we have changed variables of integration from $q_x$ to $Q_x\equiv q_xx$.
It can be shown that the integral in the last equality of (\ref{SICx1}) is equal to $-1$ for
large $x$ (specifically, for $x\gg 1/\Lambda$). Taking into account the length dependence of the coefficients  $\alpha$, $\mu_x$, and $\mu_{\rm T}^0$
as described at the beginning of this section, we get
\begin{eqnarray}
C(x,y=z=0)\propto -{\left[\ln \left(x\over\xi_{x}\right)\right]^{1\over 4}\ln\left[\ln\left( x\over\xi_{x}\right)\right]\over x^2}\,,\label{Cx}
\end{eqnarray}
where the factor $\xi_{x}^{-1}$ has been added to the arguments of log to make them dimensionless.

\vspace{.2in}

%{\cblue JT: I think we should give this calculation here. Here's a draft of it:

Another limit of the correlation function, namely $C(x=y=0,z)$, can be obtained by very similar methods. Starting with:
%\begin{eqnarray}
%C(x,y=z=0)%\nonumber \\ &&
%&=&{D\over8\pi^2\sqrt{2g\alpha}}\int_{-\infty}^{\infty} dq_x
%\int_{-\infty}^{\infty}dq_z {\sqrt{2\alpha q_z^2+gq_x^2} \exp\left(iq_xx-{q_{\perp}^2\over \Lambda^2}\right)
%\over \sqrt{\mu_xq_x^2+\mu_{\rm T}^0q_z^2}}\nonumber\\
%&=&{D\over8\pi^2\sqrt{g\mu_{\rm T}^0}}\int_{-\infty}^{\infty} dq_x\int_{-\infty}^{\infty}dq_z
%{\sqrt{q_z^2+q_2^2} \exp\left(iq_xx-{q_{\perp}^2\over \Lambda^2}\right)
%\over \sqrt{q_z^2+q_1^2}}\,,
%\label{Cx1}
%\end{eqnarray}
%where $q_1\equiv \sqrt{\mu_x/\mu_{\rm T}^0}|q_x|$, $q_2\equiv \sqrt{g/2\alpha}|q_x|$, and we have imposed a
%soft Gaussian cutoff $\Lambda$ on $q_{\perp}$. Note that the $q$-dependences
%of the coefficients $\mu_x$, $ \mu_{\rm T}^0$, $g$, and $\alpha$ imply that  $q_2\gg q_1$. %\sout{$q_2\gg q_1$, which is implied by the $q$-dependence
%%of the coefficients}.
%	
%	
%So far, we have treated our coefficients as constant when doing the inverse Fourier transform. We will now demonstrate using a different method why such an assumption is reasonable.
%		Go back to eqn. (\ref{Cx1}), and rewrite it as
\begin{eqnarray}
C(x=y=0, z)%\nonumber \\ &&
&=&{D\over8\pi^2}\int_{-\infty}^{\infty} dq_z \exp\left(iq_zz-{q_z^2\over \Lambda^2}\right)
\int_{-\infty}^{\infty}{dq_x\over\sqrt{2g\alpha}} {\sqrt{2\alpha q_z^2+gq_x^2} \exp\left(-{q_z^2\over \Lambda^2}\right)
	\over \sqrt{\mu_xq_x^2+\mu_{\rm T}^0q_z^2}}\,.
\label{Cz2_2}
\end{eqnarray}
%where all of the $q$-dependence of $g$, $\alpha$, and $\mu_{x,T}$ are kept in this expression.
We define
\begin{eqnarray}
I_z(q_z)&\equiv&%\nonumber \\ &&
\int_{-\infty}^{\infty}{dq_x\over\sqrt{2g\alpha}} {\sqrt{2\alpha q_z^2+gq_x^2} \exp\left(-{q_x^2\over \Lambda^2}\right)
	\over \sqrt{\mu_xq_x^2+\mu_{\rm T}^0q_z^2}}\,.
\label{Iqzdef}
\end{eqnarray}
%Even with the dependencies of $\bq$ in the coefficients,
One can see that this integral converges as $q_x\to0$. Hence, we now write
\beq
I_z(q_x)=I_z(q_x=0)+\delta I_z(q_x) \ ,
\label{delIzdef}
\eeq
where $\delta I_z(q_z\to0)\to0$ by definition. %{\cred [LC: Here I removed discussions about the $q_x$-dependence of the coefficients, which I suppose John only wrote to illuminate me. We have said at the begging that the calculation will be done in a harmonic approximation first, and the anharmonic effect will be taken into account at the end. Keeping those discussion may cause confusion.]}
Putting this into (\ref{Cz2_2}), we see immediately that
%, {\it even taking the $\bq$ dependence of $g$, $\alpha$, and $\mu_{x,T}$ into account},
the $I_z(q_z=0)$ piece of this gives rise {\it only} to a short-ranged contribution $\propto e^{-(\Lambda x)^2}$ to $C(x=y=0, z)$. Hence, all of the long-ranged pieces of $C(x=y=0, z)$ come from $\delta I_z(q_z)$. In the following we will first calculate ${\pp I_z(q_z)/\pp q_z^2}$, and then from that result we derive $\delta I_z(q_z)$.
%There will be two types of terms in this differentiation: those coming from differentiating the explicit $q_x$ dependences of the integrand in (\ref{Cx2}), and those that come from differentiating the implicit $q_x$ dependences of $g$, {\cmag [[CFL: As Leiming pointed out, when using the log corrections in the MT, g does not depend on $q$, right?]]} $\alpha$, and $\mu_{x,T}$. The latter, however, are always much smaller than the former. To see that, consider taking ${d\over dx} (\ln^\alpha(x) x^\beta)$. This is, of course, ${d\over dx} (\ln^\alpha(x) x^\beta)=\alpha\ln^{\alpha-1}(x)x^{\beta-1}+\beta\ln^\alpha(x) x^{\beta-1}$, where the first term comes from differentiating the $\ln$, while the second comes from differentiating the $x^\beta$. The ratio of these two terms is ${\alpha\over\beta\ln(x)}$, which clearly vanishes as $x\to0$. we can ignore any contributions to ${\pp I(q_x)/\pp q_x^2}$ coming from
%		letting the derivative act on the implicit $q_x$ dependences of $g$, $\alpha$, and $\mu_{x,T}$; that is, for the purposes of that derivative, we can replace them with constants.

Changing the variables of integration in (\ref{Iqzdef}) from $q_x$ to $Q_x$ defined via $q_x\equiv\sqrt{\mu_{\rm T}^0\over\mu_x}Q_x$, we get
\beq
{\pp \delta I_z(q_z)\over\pp q_z^2}=%{1\over\mu_{\rm T}^0}\sqrt{\mu_x\over g}\Gamma I_1(q_x)
I_3(q_z)-I_4(q_z)
\label{ppIz}
\eeq
where
\beq
I_3(q_z)\equiv {1\over g}\sqrt{\alpha\over2\mu_{\rm T}^0}\int_{-\infty}^{\infty}dQ_x { \exp\left(-{Q_x^2\over \Lambda_x^2}\right)
	\over \sqrt{(q_z^2+Q_x^2)(\Gamma^{-1} q_z^2+Q_x^2)}}\ ,
\label{I3def}
\eeq
and
\beq
I_4(q_x)\equiv {1\over2\mu_x}\sqrt{\mu_{\rm T}^0\over2\alpha}\int_{-\infty}^{\infty}dQ_x  \exp\left(-{Q_x^2\over \Lambda_x^2}\right)
\sqrt{(\Gamma^{-1} q_z^2+Q_x^2)\over(q_z^2+Q_x^2)^3}\ ,
\label{I4def}
\eeq
where we're using the same definition of $\Gamma$, namely $\Gamma\equiv{g\mu_{\rm T}^0\over 2\alpha\mu_x}$ and $\Lambda_x\equiv\Lambda\sqrt{\mu_x\over\mu^0_{\rm T}}$.

We will first calculate $I_3$. Since the integral converges rapidly before $Q_x$ is comparable to $\Lambda_x$, we can ignore the Gaussian cutoff.		
By a change of variable $Q_x =|q_z| u$, we have
\beq
I_3 \approx \frac{1}{|q_z|}{1\over g}\sqrt{2\alpha\over\mu_{\rm T}^0} \int_0^\infty \frac{d u}{\sqrt{(\Gamma^{-1} +u^2)(1+u^2)}}
\ .
\eeq
Since, as noted earlier, $\Gamma = 2g\mu_{\rm T}^0/\alpha\mu_x \sim  |\ln q|^{1/2}\to\infty$ as $q\rightarrow 0$,
we can split the integral into two parts by introducing a constant $B$ such that $\Gamma^{-1/2}\ll B\ll 1$:
\beqn
I_3 &\approx& \frac{2}{|q_z|} {1\over g}\sqrt{2\alpha\over\mu_{\rm T}^0}\left[\int_0^{B} \frac{d u}{\sqrt{\Gamma^{-1} +u^2}}+\int_{B}^\infty \frac{d u}{u\sqrt{(1+u^2)}}\right]
\\
&=& \frac{2}{|q_z|}{1\over g}\sqrt{2\alpha\over\mu_{\rm T}^0} \left[{\rm arcsinh} (\sqrt{\Gamma} B) +{\rm arcsinh} \left({1\over B}\right)\right]
\\
\label{eq:SI_I3}
&\approx&\frac{2}{|q_z|}{1\over g}\sqrt{2\alpha\over\mu_{\rm T}^0} \left[\ln (2B\sqrt{\Gamma}) +\ln \left(\frac{2}{B}\right)\right]
\\
&=&{1\over g}\sqrt{2\alpha\over\mu_{\rm T}^0}\frac{\ln (16 \Gamma)}{|q_z|}
\ ,
\eeqn
where we have used the fact that $B\gg 1$ and $\sqrt{\Gamma } \gg B$ and used the leading asymptotic expression for the ${\rm arcsinh}$ functions in (\ref{eq:SI_I3}).

We now turn to $I_4$, again with the Gaussian cutoff ignored. By the same change of variable  $Q_x =|q_z| u$, we get
\beqn
I_4(q_z)&\approx&{1\over\mu_x}\sqrt{\mu_{\rm T}^0\over2\alpha}\frac{1}{|q_z|} \int_{0}^{\infty}du
\sqrt{\Gamma^{-1} +u^2\over(1+u^2)^3}
\ .
\eeqn
The integral in this expression is easily seen to approach 1 as $\Gamma\to\infty$.
Hence,
\beqn
I_4(q_z)&\approx&{1\over\mu_x}\sqrt{\mu_{\rm T}^0\over2\alpha}\frac{1}{|q_z|}  \ .
\label{I4}
\eeqn

Comparing the expressions for $I_3$ and $I_4$, and again using the $q$ dependences of $\alpha$,  $\mu_{x}$, and $\mu_{\rm T}^0$ from equation (24) of the main text, we find that in $I_3$ the prefactor ${1\over g}\sqrt{2\alpha\over\mu_{\rm T}^0}\ln\left(16\sqrt{\Gamma}\right)\propto{\ln(|\ln(q)|)\over\sqrt{|\ln(q)|}}\to0$ as $q\to0$, while in $I_4$ the prefactor ${1\over\mu_x}\sqrt{\mu_{\rm T}^0\over2\alpha}$ is independent of $q$. Hence, we can drop $I_3$ in (\ref{ppIz}),   and obtain
\beq
{\pp \delta I_z(q_z)\over\pp q_z^2}\approx
-{1\over\mu_x}\sqrt{\mu_{\rm T}^0\over2\alpha}\frac{1}{|q_z|}
\ ,
\eeq
which  can be integrated to give:
%$I(q_x)$ by integrating $\delta I(q_x)$ with respect to $q_x^2$ to obtain
\beq
\delta I_z(q_z) \approx -{1\over\mu_x}\sqrt{2\mu_{\rm T}^0\over\alpha}|q_z|
\ .
\eeq
%Note that in the above, we have ignored the logarithmic dependence on $\bq$ in the coefficients when performing the integration for the same reason that we ignore that dependency when performing a differentiation.

Substituting this expression back into (\ref{Cz2_2}) we get
%we see that we have recovered \eq (\ref{3 regimes}), except for the constant term $\sqrt{\pi}\Lambda$, which shows does not affect the large distance behaviour.
\begin{eqnarray}
C(x,y=z=0)%\nonumber\\
&=&-{D\over 4\pi^2\mu_x}\sqrt{2\mu_{\rm T}^0\over\alpha}
\int_0^{\infty} dq_z q_z\cos\left(q_zz\right)\exp\left(-{q_z^2\over\Lambda^2}\right)\nonumber\\
&=&-{D\over 4\pi^2\mu_x}\sqrt{2\mu_{\rm T}^0\over\alpha}{1\over z^2}
\int_0^{\infty} dQ_z Q_z\cos\left(Q_z\right)\exp\left(-{Q_z^2\over\Lambda^2z^2}\right)\,,
\label{SICz1}
\end{eqnarray}
where in the second equality we have changed variables of integration from $q_z$ to $Q_z\equiv q_zz$.
The integral in the last equality of (\ref{SICz1}) is equal to $-1$ for
large $z$ (specifically, for $z\gg 1/\Lambda$). Taking into account the length dependence of the coefficients $\mu_x$, $\mu_{\rm T}^0$, and  $\alpha$
as described at the beginning of this section, we find that they  cancel out of the prefactor ${1\over\mu_x}\sqrt{\mu_{\rm T}^0\over\alpha}$. Hence, there are no logs for this direction in real space; instead we find just a simple power law:
\begin{eqnarray}
C(x=y=0, z)\propto -{1\over z^2}\,.\label{Cz}
\end{eqnarray}
%{\cred \sout{where the factor $\xi_{x}^{-1}$ has been added to the arguments of log to make them dimensionless.}}

%\sout{After essentially the same calculation we get}
%\begin{eqnarray} C(x=y=0,z)\propto {1\over z^2}\,.\end{eqnarray}

Finally let's turn to $C(x=z=0,y)$. Imposing $x=z=0$ in (\ref{SICr1}) we obtain
\begin{eqnarray}
C(x=z=0,y)
={D\over8\pi^2\sqrt{2g\alpha}}\int_{-\infty}^{\infty} dq_x\int_{-\infty}^{\infty} dq_z {\sqrt{2\alpha q_z^2+gq_x^2}
	\exp\left(-\sqrt{2\alpha q_z^2+gq_x^2}\sqrt{\mu_xq_x^2+\mu_{\rm T}^0q_z^2\over2g\alpha}|y|\right)
	\over \sqrt{\mu_xq_x^2+\mu_{\rm T}^0q_z^2}}~,
\end{eqnarray}
where we have dropped the soft cutoff at $\Lambda$  on $q_{\perp}$,  since the integral converges at
$q_{\perp}\sim 1/\sqrt{y}$, which is much smaller than $\Lambda$ for large $y$. Changing variables from
$q_{x,z}$ to $q'_{x,z}$ via $q'_x=\sqrt{\mu_x}q_x$, $q'_z=\sqrt{\mu_{\rm T}^0}q_z$, we get
\begin{eqnarray}
C(x=z=0,y)
={D\over8\pi^2\sqrt{2g\alpha\mu_x\mu_{\rm T}^0}}\int_{-\infty}^{\infty} dq'_x
\int_{-\infty}^{\infty} dq'_z {\sqrt{{2\alpha {q'_z}^2\over\mu_{\rm T}^0}+{g{q'_x}^2\over\mu_x}}
	\exp\left(-\sqrt{{2\alpha {q'_z}^2\over\mu_{\rm T}^0}+{g{q'_x}^2\over\mu_x}}\sqrt{{q'_{\perp}}^2\over2g\alpha}|y|\right)
	\over q'_{\perp}}~.
\end{eqnarray}
Switching to polar coordinates $q'_x=q'_{\perp}\cos\theta$, $q'_z=q'_{\perp}\sin\theta$, we have
\begin{eqnarray}
C(x=z=0,y)
&=&{D\over8\pi^2\sqrt{2g\alpha\mu_x\mu_{\rm T}^0}}\int_0^{2\pi} d\theta\int_0^{\infty}
dq_{\perp}\ \  {q'_{\perp}\sqrt{{2\alpha \sin^2\theta\over\mu_{\rm T}^0}+{g\cos^2\theta\over\mu_x}}} \nonumber\\
&&\times\exp\left(-\sqrt{{2\alpha \sin^2\theta\over\mu_{\rm T}^0}+{g\cos^2\theta\over\mu_x}}
\sqrt{1\over2g\alpha}{q'_{\perp}}^2|y|\right)\nonumber\\
&=&{D\over8\pi\sqrt{\mu_x\mu_{\rm T}^0}}{1\over |y|}~.
\end{eqnarray}
Taking into account the length-dependences of the coefficients we get
\begin{eqnarray}
C(x=z=0,y)\sim {\left[\ln\left(|y|\over\xi_y\right)\right]^{-{3\over 8}}\over |y|}\,.
\end{eqnarray}


\begin{thebibliography}{20}
\bibitem{footnote1}
	%\sout{\cred JT suggests moving the following (the stuff in that horrible {\cgreen green} color to a footnote:}
	The elastic energy of the system must be invariant with respect to uniform, rigid rotations about the $z$-axis. Hence strictly speaking, to meet this symmetry requirement the exact expression for the first piece in (\ref{Sli-columnar}) should be ${B\over 2}\left[\partial_y h-{(\partial_x h)^2\over 2}-{(\partial_y h)^2\over 2}\right]^2$. However, the newly added $-(\partial_y h)^2/2$ in the parenthesis only introduces anharmonic terms which are irrelevant in the long wavelength limit.
\end{thebibliography}
\end{document}